\documentclass[11pt,fleqn]{article}
\usepackage{amsfonts, epsfig, amssymb}
\usepackage{color}
\newcommand{\ol}{\setlength{\itemsep}{0pt.}\begin{enumerate}}
\newcommand{\eol}{\end{enumerate}\setlength{\itemsep}{-\parsep}}
\newcommand{\ignore}[1]{}
\title{On the entropy of a noisy function}
\author{Alex Samorodnitsky\thanks{School of Engineering and Computer Science,
The Hebrew University of Jerusalem,
Jerusalem 91904, Israel. Research supported by BSF and ISF grants.}}

\begin{document}
\date{}
\maketitle


\newtheorem{THEOREM}{Theorem}[section]
\newenvironment{theorem}{\begin{THEOREM} \hspace{-.85em} {\bf :}
}%
                        {\end{THEOREM}}
\newtheorem{LEMMA}[THEOREM]{Lemma}
\newenvironment{lemma}{\begin{LEMMA} \hspace{-.85em} {\bf :} }%
                      {\end{LEMMA}}
\newtheorem{COROLLARY}[THEOREM]{Corollary}
\newenvironment{corollary}{\begin{COROLLARY} \hspace{-.85em} {\bf
:} }%
                          {\end{COROLLARY}}
\newtheorem{PROPOSITION}[THEOREM]{Proposition}
\newenvironment{proposition}{\begin{PROPOSITION} \hspace{-.85em}
{\bf :} }%
                            {\end{PROPOSITION}}
\newtheorem{DEFINITION}[THEOREM]{Definition}
\newenvironment{definition}{\begin{DEFINITION} \hspace{-.85em} {\bf
:} \rm}%
                            {\end{DEFINITION}}
\newtheorem{EXAMPLE}[THEOREM]{Example}
\newenvironment{example}{\begin{EXAMPLE} \hspace{-.85em} {\bf :}
\rm}%
                            {\end{EXAMPLE}}
\newtheorem{CONJECTURE}[THEOREM]{Conjecture}
\newenvironment{conjecture}{\begin{CONJECTURE} \hspace{-.85em}
{\bf :} \rm}%
                            {\end{CONJECTURE}}
\newtheorem{MAINCONJECTURE}[THEOREM]{Main Conjecture}
\newenvironment{mainconjecture}{\begin{MAINCONJECTURE} \hspace{-.85em}
{\bf :} \rm}%
                            {\end{MAINCONJECTURE}}
\newtheorem{PROBLEM}[THEOREM]{Problem}
\newenvironment{problem}{\begin{PROBLEM} \hspace{-.85em} {\bf :}
\rm}%
                            {\end{PROBLEM}}
\newtheorem{QUESTION}[THEOREM]{Question}
\newenvironment{question}{\begin{QUESTION} \hspace{-.85em} {\bf :}
\rm}%
                            {\end{QUESTION}}
\newtheorem{REMARK}[THEOREM]{Remark}
\newenvironment{remark}{\begin{REMARK} \hspace{-.85em} {\bf :}
\rm}%
                            {\end{REMARK}}

\newcommand{\thm}{\begin{theorem}}
\newcommand{\lem}{\begin{lemma}}
\newcommand{\pro}{\begin{proposition}}
\newcommand{\dfn}{\begin{definition}}
\newcommand{\rem}{\begin{remark}}
\newcommand{\xam}{\begin{example}}
\newcommand{\cnj}{\begin{conjecture}}
\newcommand{\mcnj}{\begin{mainconjecture}}
\newcommand{\prb}{\begin{problem}}
\newcommand{\que}{\begin{question}}
\newcommand{\cor}{\begin{corollary}}
\newcommand{\prf}{\noindent{\bf Proof:} }
\newcommand{\ethm}{\end{theorem}}
\newcommand{\elem}{\end{lemma}}
\newcommand{\epro}{\end{proposition}}
\newcommand{\edfn}{\bbox\end{definition}}
\newcommand{\erem}{\bbox\end{remark}}
\newcommand{\exam}{\bbox\end{example}}
\newcommand{\ecnj}{\bbox\end{conjecture}}
\newcommand{\emcnj}{\bbox\end{mainconjecture}}
\newcommand{\eprb}{\bbox\end{problem}}
\newcommand{\eque}{\bbox\end{question}}
\newcommand{\ecor}{\end{corollary}}
\newcommand{\eprf}{\bbox}
\newcommand{\beqn}{\begin{equation}}
\newcommand{\eeqn}{\end{equation}}
\newcommand{\wbox}{\mbox{$\sqcap$\llap{$\sqcup$}}}
\newcommand{\bbox}{\vrule height7pt width4pt depth1pt}
\newcommand{\qed}{\bbox}
\def\sup{^}

\def\H{\{0,1\}^n}

\def\S{S(n,w)}

\def\g{g_{\ast}}
\def\xop{x_{\ast}}
\def\y{y_{\ast}}
\def\z{z_{\ast}}

\def\f{\tilde f}

\def\n{\lfloor \frac n2 \rfloor}

\def \E{\mathop{{}\mathbb E}}
\def \R{\mathbb R}
\def \Z{\mathbb Z}
\def \F{\mathbb F}
\def \T{\mathbb T}

\def \x{\textcolor{red}{x}}
\def \r{\textcolor{red}{r}}
\def \Rc{\textcolor{red}{R}}

\def \noi{{\noindent}}

\def \iff{~~~~\Leftrightarrow~~~~}

\def\<{\left<}
\def\>{\right>}
\def \({\left(}
\def \){\right)}

\def \e{\epsilon}
\def \l{\lambda}
\def \t{\tau}

\def\Tp{Tchebyshef polynomial}
\def\Tps{TchebysDeto be the maximafine $A(n,d)$ l size of a code with distance $d$hef polynomials}
\newcommand{\rarrow}{\rightarrow}

\newcommand{\larrow}{\leftarrow}

\overfullrule=0pt
\def\setof#1{\lbrace #1 \rbrace}

\begin{abstract}
\noi Let $0 < \e < 1/2$ be a noise parameter, and let $T_{\e}$ be the noise operator acting on functions on the boolean cube $\H$. Let $f$ be a nonnegative function on $\H$. We upper bound the entropy of $T_{\e} f$ by the average entropy of conditional expectations of $f$, given sets of roughly $(1-2\e)^2 \cdot n$ variables.

\noi In information-theoretic terms, we prove the following strengthening of "Mrs. Gerber's lemma": Let $X$ be a random binary vector of length $n$, and let $Z$ be a noise vector, corresponding to a binary symmetric channel with crossover probability $\e$. Then, setting $v = (1-2\e)^2 \cdot n$, we have (up to lower-order terms):
\[
H\Big(X \oplus Z\Big) \quad \ge \quad n \cdot H_2\(\e ~+~ (1-2\e) \cdot H_2^{-1}\(\frac{\E_{|B| = v} H\Big(\{X_i\}_{i\in B}\Big)}{v}\)\)
\]

\noi Assuming $\e \ge 1/2 - \delta$, for some absolute constant $\delta > 0$, this inequality, combined with a strong version of a theorem of Friedgut, Kalai, and Naor, due to Jendrej, Oleszkiewicz, and Wojtaszczyk, shows that if a boolean function $f$ is close to a characteristic function $g$ of a subcube of dimension $n-1$, then the entropy of $T_{\e} f$ is at most that of $T_{\e} g$.

\noi Taken together with a recent result of Ordentlich, Shayevitz, and Weinstein, this shows that the "Most informative boolean function" conjecture of Courtade and Kumar holds for high noise $\e \ge 1/2 - \delta$.

\noi Namely, if $X$ is uniformly distributed in $\H$ and $Y$ is obtained by flipping each coordinate of $X$ independently with probability $\e$, then, provided $\e \ge 1/2 - \delta$,  for any boolean function $f$ holds $I\(f(X);Y\) \le 1 - H(\e)$.

\end{abstract}

\section{Introduction}

\noi This paper is motivated by the following conjecture of Courtade and Kumar \cite{Kumar-Courtade}.

\noi Let $(X,Y)$ be jointly distributed in $\H$ such that their marginals are uniform and $Y$ is obtained by flipping each coordinate of $X$ independently with probability $\e$. Let $H_2$ denote the binary entropy function $H_2(x) = -x \log_2 x - (1-x) \log_2(1-x)$. The conjecture of \cite{Kumar-Courtade} is:

\cnj
\label{cnj:1}
For all boolean functions $f:~\H \rarrow \{0,1\}$,
\[
I\Big(f(X);Y\Big) \quad \le \quad 1 ~-~ H_2(\e)
\]
\ecnj

\noi This inequality holds with equality if $f$ is a characteristic function of a subcube of dimension $n-1$. Hence, the conjecture is that such functions are the "most informative" boolean functions.

\noi Following \cite{MS}, we express $I(f(X);Y)$ in terms of the 'value of the entropy functional of the image of $f$ under the noise operator' (all notions will be defined shortly). The question then becomes:

{\it Which boolean functions are the "stablest" under the action of the noise operator? That is, for which functions the entropy functional decreases the least under noise.}

\noi One can also consider a more general question of how the noise operator affects the entropy of a nonnegative function.

\noi Our main result is that for a nonnegative function $f$ on $\H$, the entropy of the image of $f$ under the noise operator with noise parameter $\e$ is upper bounded by the average entropy of conditional expectations of $f$, given sets of roughly $(1-2\e)^2 \cdot n$ variables.

\noi  As an application, using the recent strengthening \cite{JOW} of a theorem of \cite{FKN}, we show that for $\e$ close to $1/2$ characteristic functions of $(n-1)$-dimensional subcubes are at least as stable under the noise operator as functions which are close to them.

\noi This, in conjunction with \cite{FKN} and a recent result of \cite{OSW} which can be used to show that, for high noise levels $\e \sim 1/2$, boolean functions, which are potentially as stable as the characteristic functions of $(n-1)$-dimensional subcubes, have to be close to these functions, implies the validity of Conjecture~\ref{cnj:1} for high noise levels.

\subsection{Entropy of nonnegative functions and the noise operator}
We introduce some relevant notions.

\noi For a nonnegative function $f:~\H \rarrow \R$, we let the {\it entropy} of $f$ to be defined as
\[
Ent\Big(f\Big) \quad = \quad \E_x f(x) \log_2 f(x) ~~-~~ \E_x f(x) \cdot \log_2 \Big(\E f(x)\Big)
\]
We note for future use that entropy is nonnegative, homogeneous $Ent\Big(\lambda f\Big) = \lambda \cdot Ent\Big(f\Big)$ and convex in $f$ \cite{Ledoux}.

\noi Given $0 \le \e \le 1/2$, we define the {\it noise operator} acting on functions on the boolean cube as follows: for $f:~\H \rarrow \R$, we let $T_{\e} f$ at a point $x$ be the expected value of $f$ at $y$, where $y$ is $\e$-correlated with $x$. That is,
\beqn
\label{dfn:T}
\(T_{\e} f\)(x) \quad = \quad \sum_{y \in \H} \e^{|y - x|} \cdot (1-\e)^{n - |y-x|} \cdot f(y)
\eeqn
Here $|\cdot|$ denotes the Hamming distance.

\ignore{
\noi For a direction $1 \le i \le n$ we define the noise operator in direction $i$ as follows
\[
\(T_{\e_i} f\)(x) \quad = \quad \e \cdot f\(x + e_i\) ~~+~~ (1-\e) \cdot f(x)
\]
where $e_i$ is the $i^{\small{th}}$ unit vector. The operators $\left\{T_{\e_i}\right\}$ commute, and for $R \subseteq [n]$, we define $T_{\e_R}$ to be the composition of the operators $T_{\e_i}$, $i \in R$. Note that the noise operator $T_{\e}$ equals in this notation to $T_{\e_{[n]}}$.
}

\noi Note that $T_{\e} f$ is a convex combination of shifted copies of $f$. Hence, convexity of entropy implies that the noise operator decreases entropy. Our goal is to quantify this statement.

\subsubsection{Connection between notions}
\label{subsubsec:connect}

\noi Let $f$ be a nonnegative function on $\H$. Let $X$ be a random variable on $\H$ distributed according to $f/\sum f$. Let $Z$ be an independent noise random variable on $\H$. That is, $Pr\{Z = z\} ~=~ \e^{|z|} \cdot  (1-\e)^{n - |z|}$, and $X$ and $Z$ are statistically independent. Then

\begin{itemize}
\item
\[
Ent(f) \quad = \quad \E f \cdot \Big(n ~-~ H(X)\Big)
\]

\item
\[
Ent\Big(T_{\e} f\Big) \quad = \quad \E f \cdot \Big(n ~-~ H\Big(X \oplus Z\Big)\Big)
\]

\end{itemize}

\noi Let now $f:~\H \rarrow \{0,1\}$ be a boolean function, let $X$ be uniformly distributed in $\H$, let $Z$ be an independent noise random variable, and let $Y = X \oplus Z$. Then
\[
H\Big(f(X)\Big) \quad = \quad Ent\Big(f\Big) ~~+~~ Ent\Big(1-f\Big)
\]
We also have the following simple claim (proved in Section~\ref{sec:remaining} below)
\lem
\label{lem:H-entr}
For a boolean function $f:~\H \rarrow \{0,1\}$,
\[
I\Big(f(X);Y\Big) \quad = \quad  Ent\Big(T_{\e} f\Big) ~~+~~ Ent\Big(T_{\e} (1-f)\Big)
\]
\elem

\noi Therefore, Conjecture~\ref{cnj:1} translates as follows:

\cnj (An equivalent form of Conjecture~\ref{cnj:1})
\label{cnj:1-entr}

\noi For any boolean function $f:~\H \rarrow \{0,1\}$ holds
\[
Ent\Big(T_{\e} f\Big) ~~+~~ Ent\Big(T_{\e} (1-f)\Big) \quad \le \quad 1 ~-~ H_2(\e)
\]
\ecnj

\subsection{Mrs. Gerber's function and Mrs. Gerber's lemma}

\noi We describe a result from information theory, and a related function, which will be important for us \footnote{We are grateful to V. Chandar \cite{Chandar} for explaining the relevance of this result in connection to our previous work \cite{SSS} on the subject.}.

\noi Let $f_t$ be a function on the two-point space $\{0,1\}$, which is $t$ at zero and $2-t$ at one. We have
\[
Ent\Big(f_t\Big) \quad = \quad 1 ~~-~~ H_2\(\frac t2\)
\]

\noi Let $\phi(x,\e)$ be a function on $[0,1] \times [0,1/2]$ defined as follows:
\beqn
\label{mrs_gerber}
\phi(x,\e) \quad = \quad Ent\Big(T_{\e} f_t\Big)
\eeqn
where $t$ is chosen so that $Ent\(f_t\) = x$.

\noi This function was introduced in \cite{Wyner-Ziv}. We will now describe some of its properties.

\noi Note that $\phi$ is increasing in $x$, starting from zero at $x = 0$.

\noi In fact, it is easy to derive the following explicit expression for $\phi$:
\[
\phi(x,\e) \quad = \quad 1 ~-~ H_2\Big((1 - 2\e) \cdot H_2^{-1}(1-x) ~+~ \e\Big)
\]

\noi A key property of $\phi$ is its concavity.

\thm (\cite{Wyner-Ziv})
\label{thm:phi-concave}
The function $\phi(x,\e)$ is concave in $x$ for any $0 \le \e \le 1/2$.
\ethm

\noi We mention a simple corollary.

\cor
\label{cor:simple}
For all $0 \le \e \le 1/2$,
\beqn
\label{phi-lambda}
\Big(1-H_2(\e)\Big) \cdot x \quad \le \quad \phi(x,\e) \quad \le \quad (1-2\e)^2 \cdot x
\eeqn
\ecor

\prf
It's easy to check $\phi(0,\e) = 0$ and $\phi(1,\e) = 1 - H_2(\e)$. And, it's easy to check that $\frac{\partial \phi}{\partial x}$ at $x = 0$ is $(1-2\e)^2$.
\eprf

\noi From now on, when the value of $\e$ is clear from the context, we omit the second parameter in $\phi$ and write $\phi(x)$ instead of $\phi(x,\e)$.

\noi We now describe an inequality of \cite{Wyner-Ziv}, which is known as Mrs. Gerber's lemma. Following this usage, we will refer to the function $\phi$ as Mrs. Gerber's function.

\noi This inequality upperbounds the entropy of the image of a nonnegative function under the action of the noise operator. We present it in terms of the entropy functional and the noise operator\footnote{As pointed out to us by Chandar \cite{Chandar}, this is equivalent to the standard information-theoretic formulation: Let $X$ be a random binary vector of length $n$ distributed according to $f/\sum f$, and let $Z$ be a noise vector, corresponding to a binary symmetric channel with crossover probability $\e$. Then $H\(X \oplus Z\) \ge n H_2\(\e + (1-2\e) \cdot H_2^{-1}\(\frac{H(X)}{n}\)\)$.}.

\thm (\cite{Wyner-Ziv})
\label{thm:mrs-gerber}
Let $f$ be a nonnegative function on $\H$. Then
\beqn
\label{mrs-gerber-inequality}
Ent\Big(T_{\e} f\Big) \quad \le \quad n \E f \cdot \phi\(\frac{Ent(f)}{n \E f}, ~\e\)
\eeqn
\ethm

\subsection{Main results}

\noi For $A \subseteq [n]$ and for a nonnegative function $f:~\H \rarrow \R$, we denote
\[
\E\Big(f~|~A\Big) \quad = \quad \E\Big(f~\Big |~\left\{x_i\right\}_{i \in A}\Big)
\]

\noi Here $\E$ is the conditional expectation operator. That is, $\E\Big(f~|~A\Big)$ is the function of the variables $\left\{x_i\right\}_{i \in A}$, defined as the expectation of $f$ given the values of $\left\{x_i\right\}$.\footnote{We also may (and will) view $\E\(f~|~A\)$ as a function on $\H$, which depends only on variables with indices in $A$.}

\noi We write
\[
Ent\Big(f~|~A\Big) \quad = \quad Ent\bigg(\E\Big(f~|~A\Big)\bigg)
\]

\noi To connect notions, observe that if $X$ is a random variable on $\H$ distributed according to $f/\sum f$, then the distribution of $\{X_i\}_{i \in A}$ on the $|A|$-dimensional cube is given by $\frac{1}{2^{|A|} \E f} \cdot \E \(f | A\)$ and that
\beqn
\label{connect-dist}
Ent\Big(f~|~A\Big) \quad = \quad \E f \cdot \bigg(|A| ~-~ H\Big(\{X_i\}_{i \in A}\Big)\bigg)
\eeqn

\noi Our main claim is that the entropy of a nonnegative function $f$ under noise is upper bounded by the average entropy of conditional expectations of $f$, given certain random subsets of variables. We present several results which illustrate this fact.

\thm
\label{thm:main}
Let $f$ be a nonnegative function on the cube with  $\E f = 1$.

\noi Let $0 < \e < 1$ be a noise parameter. Let $T$ be a random subset of $[n]$ generated by sampling each element $i \in [n]$ independently with probability $(1-2\e)^2$. Then
\[
Ent\Big(T_{\e} f\Big) ~~\le~~ \E_T ~~\bigg(Ent\Big(f~|~T\Big) ~-~ \sum_{i \in T} Ent\Big(f~|~\{i\}\Big) \bigg) ~~+~~ \sum_{i=1}^n \phi\bigg(Ent\Big(f~ \Big |~\{i\}\Big)\bigg)
\]

\ethm

\rem
\label{rem:or+yury}
\noi We are grateful to O. Ordentlich for suggesting this formulation for the claim of this theorem, as well as for Theorem~\ref{thm:sMGL} below (in earlier versions the average on the RHS was taken over sets of a fixed cardinality $\sim (1-2\e)^2 \cdot n$, which led to more cumbersome calculations.)

\noi Let us also mention that Polyanskiy and Wu \cite{ Pol-gen} came up with a new and direct proof of the key claim, Proposition~\ref{pro:mi-noisy}, which does not rely on linear programming, and this was used by Ordentlich \cite{Or} to give direct proofs for Theorems~\ref{thm:main}~and~\ref{thm:sMGL}.
\erem

\noi Applying the inequality $\phi(x,\e) \le (1-2\e)^2 \cdot x$ (see (\ref{phi-lambda})) to the claim of the theorem, gives the following, more streamlined claim. (However, the somewhat stronger claim of the theorem is needed for the applications.)
\cor
\label{cor:streamline}
In the notation of Theorem~\ref{thm:main},
\[
Ent\Big(T_{\e} f\Big) ~~\le~~  \E_{T} Ent\Big(f~|~T\Big)
\]
\ecor

\noi Specializing to boolean functions, this implies the following claim.

\cor
\label{cor:info-app}
In the notation of Conjecture~\ref{cnj:1} and of Theorem~\ref{thm:main}, for a boolean function $f:~\H \rarrow \{0,1\}$ holds
\[
I\Big(f(X);Y\Big) \quad \le \quad \E_{T} ~I\Big(f(X);~\{X_i\}_{i \in T}\Big)
\]
\ecor

\rem
\label{rem:Or}
\noi Let $B$ be a random subset of $[n]$ generated by sampling each element $i \in [n]$ independently with probability $1-2\e$.

\noi As pointed out by Or Ordentlich \cite{Or}, it seems instructive to compare the bound in Corollary~\ref{cor:info-app} to the weaker bound
\[
I\Big(f(X);Y\Big) \quad \le \quad \E_{B} ~I\Big(f(X);~\{X_i\}_{i \in B}\Big)
\]
which can be obtained by the following information-theoretic argument.

\noi An equivalent way to obtain $Y$ from $X$ is to replace each coordinate of $X$ independently with a random bit, with probability $2\e$ .

\noi Let $S$ be the set of indices where the input bits were replaced with random bits, and let $B = S^c$.

\noi Using the chain rule of mutual information we have
\[
I\Big(f(X);Y\Big) ~~=~~ I\Big(f(X);Y,S\Big) ~-~ I\Big(f(X);S~|~Y\Big) ~~=~~ I\Big(f(X);Y~|~S\Big) ~-~ I\Big(f(X);S~|~Y\Big)
\]

\noi where the last equality follows since $I\Big(f(X);S\Big) = 0$.

\noi In particular, by non-negativity of mutual information
\[
I\Big(f(X);Y\Big) ~~\leq~~ I\Big(f(X);Y~|~S\Big) ~~=~~ \mathbb{E}_{B} ~I\Big(f(X);\{X_i\}_{i\in B}\Big)
\]
\erem

\noi We also show a somewhat different strengthening of Corollary~\ref{cor:streamline}, which gives a stronger version of Mrs. Gerber's lemma (Theorem~\ref{thm:mrs-gerber}).
\thm
\label{thm:sMGL}
In the notation of Theorem~\ref{thm:main}, setting $t = (1-2\e)^2 \cdot n$, the following is true:
\[
Ent\Big(T_{\e} f\Big) \quad \le \quad n \cdot \phi\(\frac{\E_{T} Ent\Big(f~|~T\Big)}{t}~,~~\e\)
\]
\ethm

\noi In the standard information-theoretic notation, this could be restated as follows. Let $X$ be a random binary vector of length $n$, and let $Z$ be an independent noise vector, corresponding to a binary symmetric channel with crossover probability $\e$. Then
\beqn
\label{sMGL-infor}
H\Big(X \oplus Z\Big) ~~ \ge~~ n \cdot H_2\(\e + (1-2\e) \cdot H_2^{-1}\(\frac{\E_{T} H\Big(\{X_i\}_{i\in T}\Big)}{t}\)\)
\eeqn

\noi We refer to \cite{O} for an application of (\ref{sMGL-infor}).

\rem
\label{rem:sMGL}

\noi Up to a negligible error term, the claim of the theorem is stronger than that of Theorem~\ref{thm:mrs-gerber}, since the sequence $a_t ~=~ \frac{\E_{|T| = t} H\(\{X_i\}_{i\in T}\)}{t}$ is increasing, by Han's inequality \cite{Han}.

\erem

\noi We now return to Conjectures~\ref{cnj:1}~and~\ref{cnj:1-entr}.

\noi Let us first describe a family of functions for which these conjectures are known to hold with equality. Let $1 \le k \le n$ be an index, and let $g_k(x) = 1$ if and only if $x_k = 0$. (That is, $g_k$ is a characteristic function of the $(n-1)$-dimensional subcube $\{x_k = 0\}$.)

\noi It is easy to verify that $Ent\(T_{\e} g_k\) = \frac12 \cdot \(1 - H_2(\e)\)$ and $Ent\(T_{\e} g_k\) + Ent\(T_{\e} \(1-g_k\)\) = 1 - H_2(\e)$.

\noi We  apply Theorem~\ref{thm:main} to show that, for $\e \sim 1/2$, the conjectures also hold for functions which are close to characteristic functions of subcubes.

\noi To make the notion of proximity more precise, recall (see \cite{OD}) that any function $f:\H \rarrow \R$ can be expanded in terms of the Walsh-Fourier basis:   $f(x) ~=~ \sum_{S \subseteq [n]} \widehat{f}(S) \cdot W_S(x)$. Here $W_S(x) = (-1)^{\sum_{i \in S} x_i}$.

\noi The Walsh-Fourier expansion of $g_k$ is especially simple: $\widehat{g_k}(0) = \E g_k = 1/2$, $\widehat{g_k}(\{k\}) = 1/2$, and $\widehat{g_k}(S) = 0$ for all other $S \subseteq [n]$.

\noi It follows from \cite{JOW} and \cite{FKN} that a boolean function whose Walsh-Fourier expansion is close to that of $g_k$, in that it has a large (i.e., close to $1/2$) Fourier coefficient at $\{k\}$, has to be very close, in the appropriate sense, to $g_k$.

\noi The next claim shows the conjectures to hold for such functions.

\thm
\label{thm:KC-ent-sn}
There exists an absolute constant $\delta > 0$ such that for any noise $\e \ge 0$ with $(1-2\e)^2 \le \delta$ and for any boolean function $f:~\H \rarrow \{0,1\}$ such that
\begin{itemize}
\item

$\frac12 - \delta \le \E f \le \frac12$;

\item

There exists $1 \le k \le n$ such that $|\widehat{f}(\{k\})| \ge (1 - \delta) \cdot \E f$

\end{itemize}

\noi Holds
\begin{enumerate}
\item
\[
Ent\Big(T_{\e} f\Big) \quad \le \quad \frac12 \cdot \Big(1 - H_2(\e)\Big)
\]
\item
\[
Ent\Big(T_{\e} f\Big) ~+~ Ent\Big(T_{\e} (1-f)\Big)\quad \le \quad 1 - H_2(\e)
\]

\end{enumerate}
\ethm

\noi This, in conjunction with \cite{FKN} and \cite{OSW}, which can be used to show that, for noise parameter close to $1/2$, boolean functions, which are potentially as stable as the characteristic functions of $(n-1)$-dimensional subcubes, have to satisfy the constraints of Theorem~\ref{thm:KC-ent-sn}, implies the validity of Conjecture~\ref{cnj:1} for high noise levels.

\thm
\label{thm:KC-bsn}
There exists an absolute constant $\delta > 0$ such that for any noise $\e \ge 0$ with $(1-2\e)^2 \le \delta$ and for any boolean function $f:~\H \rarrow \{0,1\}$ holds
\[
I\Big(f(X);Y\Big) \quad \le \quad 1 ~-~ H_2(\e)
\]
\ethm

\subsection{More on Theorems~\ref{thm:main} and \ref{thm:sMGL}}
\noi In this subsection we give a high-level description of the proofs of these theorems and argue that both their claims may be viewed as strengthenings of Mrs. Gerber's lemma.

\noi {\it Notation}: For a direction $1 \le i \le n$ we define the noise operator in direction $i$ as follows:
\[
\(T_{\e_{\{i\}}} f\)(x) ~~=~~ \e \cdot f\Big(x + e_i\Big) ~~+~~ (1-\e) \cdot f(x)
\]
where $e_i$ is the $i^{\small{th}}$ unit vector. The operators $\left\{T_{\e_{\{i\}}}\right\}$ commute and, for $R \subseteq [n]$, we define $T_{\e_R}$ to be the composition of  $T_{\e_{\{i\}}}$, $i \in R$. Note that the noise operator $T_{\e}$ would be written in this notation as $T_{\e_{[n]}}$.

\noi We start with the proof of Mrs. Gerber's lemma (\ref{mrs-gerber-inequality}). Since both sides of the inequality are homogeneous in $f$, we may assume $\E f = 1$.

\noi By the chain rule for entropy, for any permutation $\sigma$ in the symmetric group $S_n$ holds

\[
Ent\Big(T_{\e} f\Big) ~~=~~ \sum_{i=1}^n \Bigg(Ent\Big(T_{\e} f~|~\left\{\sigma(1),\ldots,\sigma(i)\right\}\Big) ~-~ Ent\Big(T_{\e} f~|~\left\{\sigma(1),\ldots,\sigma(i-1)\right\}\Big)\Bigg) ~~=
\]
\[
\sum_{i=1}^n \Bigg(Ent\Big(T_{\e_{\left\{\sigma(1),\ldots,\sigma(i)\right\}}} f~|~\left\{\sigma(1),\ldots,\sigma(i)\right\}\Big) ~-~ Ent\Big(T_{\e_{\left\{\sigma(1),\ldots,\sigma(i-1)\right\}}} f~|~\left\{\sigma(1),\ldots,\sigma(i-1)\right\}\Big)\Bigg) ~~\le
\]
\beqn
\label{chain-rule}
\sum_{i=1}^n \phi\Bigg( Ent\Big(T_{\e_{\left\{\sigma(1),\ldots,\sigma(i-1)\right\}}} f~|~\left\{\sigma(1),\ldots,\sigma(i)\right\}\Big) ~-~ Ent\Big(T_{\e_{\left\{\sigma(1),\ldots,\sigma(i-1)\right\}}} f~|~\left\{\sigma(1),\ldots,\sigma(i-1)\right\}\Big) \Bigg)
\eeqn

\noi Let us explain the last inequality. Let $y \in \{0,1\}^{i-1}$. Let $\tilde{f}_y$ be a function on $\{0,1\}$ defined by the restriction of the function $\E\Big(T_{\e_{\left\{\sigma(1),\ldots,\sigma(i-1)\right\}}} f~|~\{\sigma(1),\ldots,\sigma(i)\}\Big)$, which we view as a function on the $i$-dimensional cube, to the points in which the coordinates $\sigma(k)$, $k = 1,...,i-1$ are set to be $y_k$. Then, it is easy to see that
\[
Ent\Big(T_{\e_{\left\{\sigma(1),\ldots,\sigma(i)\right\}}} f~|~\left\{\sigma(1),\ldots,\sigma(i)\right\}\Big) ~-~ Ent\Big(T_{\e_{\left\{\sigma(1),\ldots,\sigma(i)\right\}}} f~|~\left\{\sigma(1),\ldots,\sigma(i-1)\right\}\Big) \quad =
\]
\[
\E_y Ent\Big(T_{\e} \tilde{f}_y\Big) \quad=\quad \E_y \Bigg(\E \tilde{f}_y \cdot \phi\(Ent\(\frac{\tilde{f}_y}{\E \tilde{f}_y}\)\)\Bigg) \quad \le \quad \phi\bigg(\E_y Ent\Big(\tilde{f}_y\Big)\bigg) \quad =
\]
\[
\phi\bigg(Ent\Big(T_{\e_{\left\{\sigma(1),\ldots,\sigma(i-1)\right\}}} f~|~\left\{\sigma(1),\ldots,\sigma(i)\right\}\Big) ~-~ Ent\Big(T_{\e_{\left\{\sigma(1),\ldots,\sigma(i-1)\right\}}} f~|~\left\{\sigma(1),\ldots,\sigma(i-1)\right\}\Big)\bigg)
\]

\noi The first equality in the second row follows from (\ref{mrs_gerber}) and the linearity of entropy. The inequality follows from concavity of the function $\phi$ and the fact that
$\E_y \E \tilde{f}_y = \E\Big(T_{\e_{\left\{\sigma(1),\ldots,\sigma(i)\right\}}} f~|~\left\{\sigma(1),\ldots,\sigma(i)\right\}\Big) \\
~=~ \E f = 1$.

\noi We now continue from (\ref{chain-rule}).

\noi For $y \in \{0,1\}^{i-1}$, let $f_y$ be a function on $\{0,1\}$ defined by the restriction of the function $\E\Big( f~|~\{\sigma(1),\ldots,\sigma(i)\}\Big)$ to the points in which the coordinates $\sigma(k)$, $k = 1,...,i-1$ are set to be $y_k$.

\noi Since the noise operator $T_{\e_{\left\{\sigma(1),\ldots,\sigma(i-1)\right\}}}$ is stochastic, the functions $\Big\{\tilde{f}_y\Big\}$ are a stochastic mixture of the functions $\Big\{f_y\Big\}$. Hence, since the $Ent$ functional is convex, for any $0 \le \e \le 1$ holds
\[
Ent\Big(T_{\e_{\left\{\sigma(1),\ldots,\sigma(i-1)\right\}}} f~|~\left\{\sigma(1),\ldots,\sigma(i)\right\}\Big) ~-~ Ent\Big(T_{\e_{\left\{\sigma(1),\ldots,\sigma(i-1)\right\}}} f~|~\left\{\sigma(1),\ldots,\sigma(i-1)\right\}\Big) ~=
\]
\beqn
\label{noisy-mi}
\E_y Ent\Big(\tilde{f}_y\Big) \quad \le \quad  \E_y Ent\Big(f_y\Big) \quad =
\eeqn
\[
Ent\Big(f~|~\left\{\sigma(1),\ldots,\sigma(i)\right\}\Big) ~-~ Ent\Big(f~|~\left\{\sigma(1),\ldots,\sigma(i-1)\right\}\Big)
\]

\noi And hence (\ref{chain-rule}) is upper bounded by
\[
\sum_{i=1}^n \phi\bigg(Ent\Big(f~|~\left\{\sigma(1),\ldots,\sigma(i)\right\}\Big) ~-~ Ent\Big(f~|~\left\{\sigma(1),\ldots,\sigma(i-1)\right\}\Big)\bigg) ~\le~ n \cdot \phi\(\frac{Ent\Big(f\Big)}{n}\)
\]
\noi where in the last inequality the concavity of $\phi$ is used again.

\subsubsection{Our improvement}
\label{subsubsec:improve}

\noi We attempt to quantify the loss in inequality (\ref{noisy-mi}).

\noi Let us introduce some notation. For a nonnegative function $g$ on the cube, for a subset $A \subset [n]$, and for an element $m \not \in A$, we define

\[
I_g(A,m) \quad = \quad Ent\Big(g~|~A \cup \left\{m\right\}\Big) ~~-~~ Ent\Big(g~|~A\Big) ~~-~Ent\Big(g~|~\left\{m\right\}\Big)
\]

\noi This quantity is always nonnegative. In fact, let $X$ be distributed on $\H$ according to $g/\sum g$. Assume $\E g = 1$ and note that in this case, by Subsection~\ref{subsubsec:connect} and by (\ref{connect-dist}), we have $I_g(A,m) ~=~ H\(\{X_i\}_{i \in A}\) + H\(X_m\) - H\(\{X_j\}_{j \in A \cup \{m\}}\) ~=~ I\(\{X_i\}_{i \in A};~X_m\)$.

\noi Coming back to (\ref{noisy-mi}), observe that
$Ent\Big(T_{\e_{\left\{\sigma(1),\ldots,\sigma(i-1)\right\}}} f~|~\left\{\sigma(i)\right\}\Big) ~=~ Ent\Big( f~|~\left\{\sigma(i)\right\}\Big)$.

\noi Hence, taking $A = \left\{\sigma(1),\ldots,\sigma(i-1)\right\}$ and $m = \left\{\sigma(i)\right\}$, the decrease in (\ref{noisy-mi}) is from $I_f(A,m)$ to $I_{T_{\e_A} f}(A,m)$. Therefore, our goal is to quantify the decrease in mutual information in the presence of noise.

\noi In the next two sections we consider a somewhat more general question of upper bounding $I_{T_{\e_A} f}(A,m)$, given $f$, $A$, and $m$. In Section~\ref{sec:noisy-mi} we upper bound $I_{T_{\e_A} f}(A,m)$ by the value of a certain linear program. In Section~\ref{sec:sym} we introduce a symmetric version of this program and a symmetric solution for the symmetric program, and show its value to be at least as large as that of the original program.

\noi We then find the value of the symmetric solution, as a function of $f$, $A$, and $m$. This value provides an upper bound on the noisy mutual information (see Proposition~\ref{pro:mi-noisy}).

\noi In order to prove Theorems~\ref{thm:main}~and~\ref{thm:sMGL} we apply the improved bound in (\ref{noisy-mi}), averaging the chain rule for the entropy of $T_{\e} f$ over all permutations $\sigma \in S_n$.

\noi This improvement in (\ref{noisy-mi}) is the reason we suggest to view both these claims as stronger versions of Mrs. Gerber's lemma.

\noi On the other hand, strictly speaking, this line of argument does not necessarily provide a direct improvement of (\ref{mrs-gerber-inequality}), since in the averaging step we have to replace $\phi(x,\e)$ by a larger linear function $(1-2\e)^2 \cdot x$, in order to be able to come up with manageable estimates.

\noi In fact, the difference between the two claims stems from the different ways in which we apply this "linearization" of the function $\phi(x,\e)$ during averaging. The bounds they give are incomparable, though Theorem~\ref{thm:sMGL} is a more evident improvement of (\ref{mrs-gerber-inequality}).

\noi We note that the two functions $\phi(x,\e)$ and $(1-2\e)^2 \cdot x$ almost coincide for small values of $x$, and, loosely speaking, if the entropy of $f$ is not too large, as is the case, say, for boolean functions, all the arguments of $\phi$ should lie very close to zero, meaning not much lost in the linear approximation. In this case, the bounds in Theorems~\ref{thm:main}~and~\ref{thm:sMGL} are very close to that in Corollary~\ref{cor:streamline}.

\subsubsection{Related work}

\noi Y. Polyanskiy \cite{Polyanskiy} has pointed out to us that the related question of upper bounding $I_{T_{\e_A} f}(A,m)$ given $I_f(A,m)$ belongs to the area of {\it strong data processing inequalities} (SDPI) in information theory (see \cite{Pol-Gaus}, \cite{Pol-gen} for pertinent results, and, in particular, for a new proof of Proposition~\ref{pro:mi-noisy}).

\subsubsection*{Organization of the paper}

\noi This paper is organized as follows. The proof of Theorem~\ref{thm:main} is given in Sections~\ref{sec:noisy-mi}~to~\ref{sec:proof-average}. Theorem~\ref{thm:KC-ent-sn} is proved in Section~\ref{sec:kc}. The remaining proofs are presented in Section~\ref{sec:remaining}.

\section{A linear programming bound for noisy mutual information}
\label{sec:noisy-mi}

\noi In this section we upper bound the noisy mutual information $I_{T_{\e_A} f}(A,m)$ by the value of a certain linear program.

\noi Let $f$ be a nonnegative function on the cube. Let $A$ be a subset of $[n]$ and let $m \not \in A$.

\noi Let $|A| = k$. We will assume, without loss of generality, that $A = [k]$ and that $m = k+1$.

\noi {\it Notation:} From now on, we write $\l$ for $(1-2\e)^2$.

\subsubsection*{Discussion}

\noi Before going into details, let us give a high-level description of what the linear program attempts to capture. For ease of discussion the notation we use here is slightly different from that in the definition of the program below (they are the same up to scaling).

\noi Given a random variable $X$ on $\H$ distributed according to $f / \sum f$, consider a function $I$ on the $k$-dimensional boolean cube, defined for $S \subseteq [k]$ by the mutual information $I(S) ~=~ I(\{X_i\}_{i \in S};~X_{k+1})$.

\noi For $S \subseteq [k]$ and for $i \in S$, let $y_{S,i} = I(S) - I(S \setminus \{i\})$ be the "discrete derivative" of $I$ at $S$ in direction $i$. Note that $y_{S,i} \ge 0$, since this is the mutual information between $X_i$ and $X_{k+1}$, given $\{X_j\}_{j \in S \setminus \{i\}}$. We view $y$ as a function on the edges of the cube. Note also that, for any $S$, the value of the summation of $y$ on the edges of any path from $\emptyset$ to $S$ is $I(S)$.

\noi For $R \subseteq [k]$, applying noise in directions in $R$ to $f$ leads to a new distribution $T_{\e_R} f/ \(\sum T_{\e_R} f\)$ on $\H$. This defines a new random variable $X^R$, a mutual information function $I^R$ and discrete derivative functions $x^R_{S,i} = I^R(S) - I^R(S \setminus \{i\})$. (Note that $x^{\emptyset} = y$).

\noi Observe that noise decreases mutual information, and hence $I^R \le I$. However, the discrete derivatives $x^R$ do not necessarily decrease. With that, and this is a key fact, by the {\it strong data processing inequality} \cite{AG}, noise in direction $i$ decreases the discrete derivative in direction $i$ (i.e., the conditional mutual information between $X^R_i$ and $X^R_{k+1}$) by a factor of at least $\l$.

\noi The variables in the linear program below are the values of the discrete derivates $x^R$, while we consider the discrete derivatives $y = x^{\emptyset}$ related to the initial function $f$ to be the boundary data of the program. We note that the noisy mutual information $I([k]) = I_{T_{\e_{[k]}} f}([k],k+1)$ is a linear combination of the variables, and that the strong data processing inequality provides linear local constraints on the variables.

\noi Finally, we would like to explain the intuition behind the symmetrization procedure in Section~\ref{sec:sym}. The fact that for any $R$ and $S$ the value of the summation of $x^R$ on the edges of any path from $\emptyset$ to $S$ is $I^R(S)$ provides a family of "symmetric" linear constraints on the variables. This makes it natural to look for a symmetric feasible solution to the linear program (symmetrizing the boundary data accordingly), one in which $x^R(S,i)$ depends only on $|S|$ and on $|R \cap S|$.

\noi We were led to expect that this symmetric solution would be an optimal one by the following informal speculation. It turns out that the strong data processing inequality $x^{R}_{S,i} \le \l \cdot \(x^{R \setminus i}_{S,i}\)$ may be replaced by a stronger inequality $x^{R}_{S,i} \le \phi\(x^{R \setminus i}_{S,i}\)$ (see (\ref{phi-lambda})).\footnote{This was shown in \cite{S-manu} if $f$ is monotone (which suffices for applications) and in \cite{Pol-gen} for general functions.} This turns the program into a {\it strictly concave} optimization problem, for which optimality of a feasible symmetric solution might be anticipated. It might also be hoped for that replacing the concave constraint by a linear one would preserve this property, and this is indeed turns out to be true.

\noi More to the point, it turns out that for the symmetric solution we define, all the inequalities $x^{R}_{S,i} \le \l \cdot \(x^{R \setminus i}_{S,i}\)$ hold with equality.

\noi The resulting argument is straightforward, most of the work going into setting up notation, and verifying feasibility of the symmetric solution. The key step, relying on symmetric properties of the discrete cube, is made in Lemma~\ref{lem:simple-non-sym}.


\subsubsection*{Linear program}

\noi {\it Boundary data}: For $S \subseteq [k]$ and for $i \in S$, we write
\[
y_{S,i} ~~=~~ Ent\Big(f~|~S \cup \{k+1\}\Big) ~-~ Ent\Big(f~|~S \setminus \{i\} \cup \{k+1\} \Big) ~-~ Ent\Big(f~|~S\Big) ~+~ Ent\Big(f~|~S \setminus \{i\}\Big)
\]
The numbers $\{y_{S,i}\}$ are the boundary data for this problem.

\noi We note that $y_{S,i} \ge 0$ for all $S$ and $i$.  In fact, the value of $y_{S,i}$ is proportional to a certain conditional mutual information. To see this, let $X$ be distributed on $\H$ according to $f/\sum f$. Assume $\E f = 1$ and note that, by Subsection~\ref{subsubsec:connect} and by (\ref{connect-dist}), $y_{S,i}$ is given by  \\
$H\(\{X_i\}_{i \in S \setminus \{i\} \cup \{k+1\}}\) + H\(\{X_i\}_{i \in S}\) - H\(\{X_i\}_{i \in S \cup \{k+1\}}\) - H\(\{X_i\}_{i \in S \setminus \{i\}}\) ~=~ I\(X_i;X_{k+1}|\{X_j\}_{j \in S \setminus \{i\}}\)$.

\noi {\it Variables}: $x^R_{S,i}$ for $R,S \subseteq [k]$ and $i \in S$.

\noi {\it The optimization problem:} Given the boundary data, we want to upper bound $\mu$, where
\beqn
\label{OP}
\mu \quad = \quad \mbox{Max} \quad \sum_{i=1}^{k} x^{[k]}_{\{1,...,i\};~i}
\eeqn
under the following constraints.

\noi {\bf Constraints}:
\begin{enumerate}
\item
\[
x^{\emptyset}_{S,i} \quad = \quad y_{S,i}
\]

\item
\[
x^{R}_{S,i} \quad = \quad x^{R \cap S}_{S,i}
\]

\item
For all $\sigma, \tau \in S_k$ holds
\[
\sum_{i=1}^{k} x^{R}_{\{\sigma(1),...,\sigma(i)\},~\sigma(i)} \quad = \quad \sum_{i=1}^{k} x^{R}_{\{\tau(1),...,\tau(i)\},~\tau(i)}
\]

\item
If $i \in R$ then
\[
x^{R}_{S,i} \quad \le \quad \l \cdot \Big(x^{R \setminus i}_{S,i}\Big)
\]

\end{enumerate}

\noi We then have the following claim.
\thm
\label{thm:mi_opt}
The noisy mutual information $I_{T_{\e_{[k]}} f}\Big([k],k+1\Big)$ is upperbounded by the value of the optimization problem (\ref{OP}).
\ethm
\prf

\noi First, consider the boundary data.
\ignore{Note that for all $S$ and $i$ we have $y_{S,i} \ge 0$. This follows easily from supermodularity of the entropy functional \footnote{In fact, $y_{S,i}$ is a known notion in information theory. It is the {\it mutual information between $m$ and $k+1$, given $S \setminus \{m\}$}.}.
}
\ignore{
For $S \subseteq [k]$ and for $i \in S$, we write
\[
y_{S,i} ~~=~~ Ent\Big(f~|~S \cup \{k+1\}\Big) ~-~ Ent\Big(f~|~S \setminus \{i\} \cup \{k+1\} \Big) ~-~ Ent\Big(f~|~S\Big) ~+~ Ent\Big(f~|~S \setminus \{i\}\Big)
\]
}
We claim that for any permutation $\sigma \in S_k$ holds
\beqn
\label{mi-chain}
\sum_{i=1}^{k} y_{\{\sigma(1),...,\sigma(i)\},~\sigma(i)} \quad = \quad I_f\Big([k],k+1\Big)
\eeqn
In fact, it is easy to see that the LHS is a telescopic sum, summing to
\[
Ent\Big(f~|~[k+1]\Big) ~~-~~ Ent\Big(f~|~[k]\Big) ~~-~~ Ent\Big(f~|~\{k+1\}\Big) \quad = \quad I_f\Big([k],k+1\Big)
\]

\noi Next we define a feasible solution for (\ref{OP}) whose value is $I_{T_{\e_{[k]}} f}\Big([k],k+1\Big)$.

\noi Fix $R \subseteq [k]$. Write $f^{R}$ for $T_{\e_R} f$. For $S \subseteq [k]$ and $i \in S$  set

\[
x^R_{S,i} ~=~ Ent\Big(f^R~|~S \cup \{k+1\}\Big) - Ent\Big(f^R~|~S \setminus \{i\} \cup \{k+1\} \Big) - Ent\Big(f^R~|~S\Big) + Ent\Big(f^R~|~S \setminus \{i\}\Big)
\]

\noi Clearly, $x^{\emptyset}_{S,i} = y_{S,i}$ and hence the first constraint of the program is satisfied.

\noi As above, for any permutation $\sigma \in S_k$ holds
\[
\sum_{i=1}^{k} x^R_{\{\sigma(1),...,\sigma(i)\},~\sigma(i)} \quad = \quad I_{T_{\e_R}} f\Big([k],k+1\Big)
\]

\noi Hence, the third constraint is satisfied as well.

\noi In particular,
\[
\sum_{i=1}^{k} x^{[k]}_{\{1,...,i\},~i} \quad = \quad  I_{T_{\e_{[k]}}} f\Big([k],k+1\Big)
\]

\noi so,  the value given by this solution is indeed $I_{T_{\e_{[k]}}} f\Big([k],k+1\Big)$.

\noi We continue to prove its feasibility. We claim that for any $A \subseteq [k]$ holds $Ent\Big(f^R~|~A\Big) ~=~ Ent\Big(f^{R \cap A}~|~A\Big)$.

\noi To see this, note that the noise operators commute with the conditional expectation operators, and hence
\[
\E\Big(T_{\e_R} f~|~A\Big) ~=~ T_{\e_R} \E\Big(f~|~A\Big) ~=~  T_{\e_{R \cap A}} T_{\e_{R \setminus A}} \E\Big(f~|~A\Big) ~=~ T_{\e_{R \cap A}} \E\Big(f~|~A\Big) ~=~ \E\Big(T_{\e_{R \cap A}} f~|~A\Big)
\]

\noi Hence, by definition, $x^R_{S,i} = x^{R \cap S}_{S,i}$ for any $R, S \subseteq [k]$, and the second constraint holds.

\noi To conclude the proof of the theorem, it remains to show that for any $R \subseteq S \subseteq [k]$ and $i \in R$ holds
\beqn
\label{sdpi}
x^{R}_{S,i} \quad \le \quad  \l \cdot \Big(x^{R \setminus i}_{S,i}\Big)
\eeqn

\noi Recall that the strong data processing inequality \cite{AG} for a binary symmetric channel with crossover probability $\epsilon$ states that if $V$ is  a random variable with values in $\{0,1\}$, and $U$ is any random variable; and if $Y = V \oplus Z$, where $Z$ is a Bernoulli random variable with parameter $\e$, statistically independent of $U$ and $V$, then $I(U;Y)\le \l \cdot I(U;V)$.

\noi Let $X$ be distributed on $\H$ according to $f^{R \setminus \{i\}}/\sum f^{R \setminus \{i\}}$. Assuming, as we may, $\E f = \E f^{R \setminus \{i\}} = 1$, we can rewrite (\ref{sdpi}) as
\[
I\(X_i \oplus Z~;X_{k+1}~\Big |~ \Big\{X_j\Big\}_{j\in S \setminus \{i\}}\) ~~\le~~ \l \cdot I\(X_i ~;X_{k+1}~\Big |~ \Big\{X_j\Big\}_{j\in S \setminus \{i\}}\)
\]
which follows from applying the strong data processing inequality with $U = X_{k+1}$ and $V = X_i$, both conditioned on $\{X_j=x_j\}_{j\in S \setminus \{i\}}$, for all values of ${x_j}$.

\section{The optimization problem and its symmetric version}
\label{sec:sym}

In this section we introduce a symmetric version of the optimization problem (\ref{OP}) and a specific symmetric feasible solution for the symmetric problem. We then argue that the value of this solution for the symmetric problem is at least as large as the optimal value for the original problem. Hence this value provides an upper bound on the noisy mutual information.

\subsection{The symmetric problem and solution}
\label{subsec:symm-feasible}

\noi Let $\left\{x^R_{S,i}\right\}$ be a feasible solution to the optimization problem (\ref{OP}) with boundary data $\left\{y_{S,i}\right\}$.

\noi We define numbers $y_1,\ldots,y_k$ as follows. For $1 \le s \le k$ let
\beqn
\label{y_S}
y_s  \quad = \quad \E_{(S,i)} ~y_{S,i}
\eeqn
where the expectation is taken over all pairs $(S,i)$ such that $|S| = s$ and $i \in S$.

\noi For $0 \le r < s \le k$ we define $x^r_s$ recursively in the following manner:
\beqn
\label{x_S}
x^r_s \quad = \quad \left\{\begin{array}{lll} y_s & \mbox{if} & r = 0 \\  \l \cdot x^{r-1}_s ~+~ (1-\l) \cdot x^{r-1}_{s-1} &  \mbox{otherwise} & \end{array} \right.
\eeqn

\noi We now define the {\it symmetric version} of (\ref{OP}), by replacing the boundary data by a new, symmetric one. We set, for all $i \in S \subseteq [k]$ with $|S| = s$:
\[
\bar{y}_{S,i} \quad = \quad y_{s}
\]

\noi Next, we define the {\it symmetric solution} for the symmetric problem, in the following way. For $R \subseteq S$ with $|R| = r$, we set
\[
\bar{x}^R_{S,i} \quad = \quad \left\{\begin{array}{lll} \l \cdot x^{r-1}_s & \mbox{if} & i \in R \\ x^r_s & \mbox{otherwise} & \end{array} \right.
\]
and for general $R, S$ we set
\[
\bar{x}^R_{S,i} \quad = \quad \bar{x}^{R \cap S}_{S,i}
\]

\pro
\label{pro:sym-legal}
The solution above is a feasible solution of the symmetric version of (\ref{OP}).

\noi Moreover, for any $R \subseteq [k]$ of cardinality $r$ and for any $\tau \in S_k$ holds
\beqn
\label{value-sym}
\sum_{i=1}^k \bar{x}^R_{\left\{\tau(1),...,\tau(i)\right\},\tau(i)} \quad = \quad \sum_{j=1}^{k - r} y_j \quad + \quad \l \cdot \sum_{t=0}^{r-1} x^t_{k-r+t+1}
\eeqn
\epro

\prf

\noi The constraints 1 and 2 of (\ref{OP}) hold, by the definition of $\bar{x}^R_{S,i}$. We pass to constraint 4. Clearly, because of constraint 2, it suffices to prove it for $R \subseteq S$. In this case, taking $i \in R$, we have, by the definition of $\bar{x}^R_{S,i}$

\[
\bar{x}^R_{S,i} ~~=~~ \l \cdot x^{r-1}_s ~~ = ~~ \l \cdot \bar{x}^{R \setminus \{i\}}_{S,i}
\]

\noi Next, we note that (\ref{value-sym}) will imply validity of constraint 3, since the RHS of (\ref{value-sym}) does not depend on $\tau$.

\noi It remains to prove (\ref{value-sym}). Let $i_1 < i_2 < ... < i_r$ be such that $R = \left\{\tau\(i_1\), \tau\(i_2\), ... , \tau\(i_r\)\right\}$. Then
\[
\sum_{i=1}^k \bar{x}^R_{\left\{\tau(1),...,\tau(i)\right\},\tau(i)} \quad = \quad \sum_{j=1}^{i_1-1} ~~+~~ \sum_{j=i_1}^{i_2 - 1} ~~+~~ \ldots ~~+~~ \sum_{j=i_r}^k \quad = \quad
\]
\[
\sum_{j=1}^{i_1-1} y_j \quad + \quad \(\l \cdot y_{i_1} + \sum_{j=i_1+1}^{i_2-1} x^1_{j}\) \quad + \quad \(\l \cdot x^1_{i_2} + \sum_{j=i_2+1}^{i_3-1} x^2_j\) \quad + \ldots \(\l \cdot x^{r-1}_{i_r} + \sum_{j=i_r+1}^k x^r_j\)
\]
Expanding $x^t_s = \l \cdot x^{t-1}_s + (1-\l) \cdot x^{t-1}_{s-1}$, we have the following exchange rule:

\noi Two adjacent summands of the form $\l \cdot x^t_j + x^{t+1}_{j+1}$ can always be replaced by $x^t_j + \l \cdot x^t_{j+1}$. Applying this appropriate number of times in each bracket transforms the expression above into

\[
\sum_{j=1}^{i_1-1} y_j \quad + \quad \(\sum_{j=i_1}^{i_2-2} y_j + \l \cdot y_{i_2-1}\) \quad + \quad \(\sum_{j=i_2}^{i_3-2} x^1_j + \l \cdot x^1_{i_3-1}\) \quad + \ldots \(\sum_{j=i_r}^{k-1} x^{r-1}_j + \l \cdot x^{r-1}_k\)
\]

\noi Next we observe that the following rules apply in the original ordering of the summands: To the right of $x^t_j$ is always either $x^t_{j+1}$ or $\l \cdot x^t_{j+1}$. To the right of $\l \cdot x^r_s$ is always either $x^{r+1}_{s+1}$ or $\l \cdot x^{r+1}_{s+1}$.

\noi Moreover, this is easily verified to be preserved by the exchange rule above, by checking the four arising cases.

\noi This means that applying the exchange rule as many times as needed, we can ensure all the summands multiplied by $\l$ to be on the last $r$ places on the right. Since the first summand is always either $y_1$ or $\l \cdot y_1$, these invariants guarantee that by doing so we obtain (\ref{value-sym}).

\eprf

\subsection{Optimality of the symmetric solution}

\thm
\label{thm:symm-opt}
Let $\left\{x^R_{S,i}\right\}$ be a feasible solution to the linear optimization problem (\ref{OP}). Let $\left\{\bar{x}^R_{S,i}\right\}$ be the symmetric solution for the symmetric version of this problem.

\noi Then, for any $0 \le r \le k$ holds:
\[
\E_{|R| = r} ~~\sum_{i=1}^k x^R_{\left\{1,...,i\right\},i} \quad \le \quad \E_{|R| = r} ~~\sum_{i=1}^k \bar{x}^R_{\left\{1,...,i\right\},i}
\]
\ethm

\cor
\label{cor:symm-opt}
The optimal value of (\ref{OP}) is upper bounded by the value of the symmetric solution to the symmetric version of the problem, which is given by
\[
\l \cdot \sum_{t=0}^{k-1} x^t_{t+1}
\]
\ecor
\prf
Apply the theorem with $r = k$ and use (\ref{value-sym}).
\eprf

\prf (Of the theorem).

\noi We proceed by double induction - on $k$ and on $0 \le r \le k$. For $k = 1$ the claim is easily seen to be true.

\noi Note also that the claim is true for any $k$ and $r = 0$. This follows from constraints 1 and 3 of the linear program (\ref{OP}) and the definition of the symmetric boundary data. In fact, we have
\[
\sum_{j=1}^k y_{\left\{1,...,j\right\},j} \quad = \quad \E_{\sigma \in S_k} ~\sum_{j=1}^{k} y_{\{\sigma(1),...,\sigma(j)\},~\sigma(j)} \quad = \quad \sum_{j=1}^{k} \E_{\sigma \in S_k} ~y_{\{\sigma(1),...,\sigma(j)\},~\sigma(j)} \quad =
\]
\[
\sum_{j=1}^{k} \E_{|S|=j,~i \in S} ~~y_{S,i} \quad = \quad \sum_{j=1}^{k} y_j \quad = \quad \sum_{j=1}^{k} \bar{y}_{\left\{1,...,j\right\},j}
\]

\noi Let now numbers $r$ and $k$, with $0 < r \le k$ be given. Assume the claim holds for $k-1$, and also for $k$, for all $0 \le t \le r-1$. We will argue it also holds for $k$ and $r$.

\noi We start with some simple properties of the linear program (\ref{OP}). We assume to be given the boundary data and a specific feasible solution to (\ref{OP}), and the symmetric solution to the symmetric version of (\ref{OP}), as in Theorem~\ref{thm:symm-opt}.

\lem
\label{lem:sol-restrict}
Let $M \subseteq [k]$. Let $\Big\{y_{K,i}\Big\}_{i \in K \subseteq M}$ be the restriction of the boundary data to subsets of $M$. For $R \subseteq M$, let $\left\{x^R_{K,i}\right\}_{i \in K \subseteq M}$ be the restriction of the feasible solution to subsets of $M$.

\noi Then $\left\{x^R_{K,i}\right\}_{i \in K \subseteq M}$ is a feasible solution to the appropriate (smaller) optimization problem on $M$.
\elem
\prf

\noi Constraints 1, 2, and 4 are easy to check. As for constraint 3, let $\sigma$ and $\tau$ be two permutations from $M$ to itself. Extend them in the same way to permutations $\sigma'$ and $\tau'$ on $[k]$. It is then easy to see that constraint 3 holds for $\sigma$ and $\tau$ in the smaller problem, since it holds for $\sigma'$ and $\tau'$ in the larger one.

\eprf

\lem
\label{lem:sym-same}
Let $M \subseteq [k]$, with $|M| = m$ and let $R \subseteq [k]$. Let $\tau$ be a bijection from $[m]$ to $M$. Let
\[
F\Big(M, R, \tau\Big) \quad = \quad \sum_{j=1}^{m} \bar{x}^R_{\left\{\tau(1),...,\tau(j)\right\},~\tau(j)}
\]

\noi Then $F\Big(M, R, \tau\Big)$ depends only on $m$ and $|R\cap M|$.
\elem

\prf

\noi Since the symmetric solution $\Big\{\bar{x}^R_{S,i}\Big\}$ satisfies constraint 2 of (\ref{OP}), we have
\[
F\Big(M, R, \tau\Big) \quad = \quad \sum_{j=1}^{m} \bar{x}^R_{\left\{\tau(1),...,\tau(j)\right\},~\tau(j)} \quad = \quad \sum_{j=1}^{m} \bar{x}^{R \cap M}_{\left\{\tau(1),...,\tau(j)\right\},~\tau(j)} \quad = \quad F\Big(M, R \cap M, \tau\Big)
\]

\noi Let $r = |R \cap M|$.

\noi Proceeding exactly as in the proof of Proposition~\ref{pro:sym-legal}, we get that
\[
F\Big(M, R \cap M, \tau\Big) \quad = \quad \sum_{j=1}^{m - r} y_j \quad + \quad \l \cdot \sum_{t=0}^{r-1} x^t_{m-r+t+1}
\]

\noi That is, $F\Big(M, R, \tau\Big)$ depends only on $m$ and $r = |R \cap M|$, as claimed.
\eprf

\noi Next, we introduce some notation.

\subsubsection{Notation}

\begin{enumerate}

\item

\noi Let $M \subseteq [k]$. Let $\Big\{y_{K,i}\Big\}_{i \in K \subseteq M}$ be the restriction of the boundary data to the subsets of $M$.

\noi We will denote by $\left\{{\cal S}_M\Big[x^R_{K,i}\Big]\right\}$ the symmetric solution to the symmetric version of the smaller problem with this boundary data.

\item

\noi Let $L \subseteq [k]$, with $L = \left\{i_1,...,i_{\ell}\right\}$, so that $i_1 < i_2 < ... < i_{\ell}$. Let $R \subseteq [k]$. Write
\[
\mu^R(L) \quad = \quad  \sum_{j=1}^{\ell} x^R_{\left\{i_1,...,i_j\right\},~i_j}
\]

\noi For $L \subseteq M \subseteq [k]$, and $R \subseteq M$, we denote
\[
{\cal S}[\mu]^R_M(L) \quad = \quad \sum_{j=1}^{\ell} {\cal S}_M\Big[x^R_{\left\{i_1,\ldots,i_j\right\},~i_j}\Big]
\]
Note that this quantity depends on $M$. With that, by Lemmas~\ref{lem:sol-restrict}~and~\ref{lem:sym-same}, given $M$, it depends only on the cardinalities $|L|$ and $|R \cap L|$.

\item

\noi Using the observation in the preceding paragraph, given $R \subseteq L \subseteq M \subseteq [k]$, with $|L| = \ell$, and $|R| = r$, we may also write ${\cal S}[\mu]^r_M\Big(\ell\Big)$ for ${\cal S}[\mu]^R_M(L)$.

\noi In particular, note that the proof of Lemma~\ref{lem:sym-same} gives, in this notation
\beqn
\label{full-sym-k-1}
{\cal S}[\mu]^r_{[k]}\Big(m\Big) \quad = \quad \sum_{j=1}^{m - r} y_j \quad + \quad \l \cdot \sum_{t=0}^{r-1} x^t_{m-r+t+1}
\eeqn

\item

\noi Finally, for $M \subseteq [k]$ and $0 \le r \le |M|$, we write
\[
\mu^r_M ~~=~~~ \E_{|R| = r,R \subseteq M} ~~ \mu^R(M) \quad \mbox{and} \quad {\cal S}[\mu]^r_M ~~=~~ \E_{|R| = r,R \subseteq M} ~~ {\cal S}[\mu]^R_M(M)
\]

\ignore{
Let $R \subseteq S \subseteq [k]$. Let $|S| = s$ and $|R| = r$. Write $S = \left\{i_1,...,i_s\right\}$. Given initial data $\left\{y_{T,i}\right\}$, for $i \in T \subseteq S$,  and a corresponding feasible solution $\left\{x^R_{T,i}\right\}$, we denote
\[
\mu^R_S = \sum_{j=1}^{s} x^R_{\left\{i_1,...,i_j\right\},~i_j} \quad \mbox{and} \quad  \bar{\mu}^R_S = \sum_{j=1}^{s} \bar{x}^{R,S}_{\left\{i_1,...,i_j\right\},~i_j}
\]
where in this case $\bar{x}^{R,S}$ stands for the symmetric solution for the symmetric data, {\it symmetrized over $S$}.

\noi Let
\[
\mu^r_S = \E_{R \subseteq S, |R| = r} ~~\mu^R_{S} \quad \mbox{and} \quad \bar{\mu}^r_S = \E_{R \subseteq S, |R| = r} ~~\bar{\mu}^R_S
\]
}

\end{enumerate}

\noi We have completed introducing the new notation. In this notation the claim of the theorem amounts to:
\beqn
\label{gen_ineq}
\mu^r_{[k]} \quad \le \quad {\cal S}[\mu]^r_{[k]}
\eeqn

\noi  We start with a lemma connecting the value of a solution of the optimization problem to these of smaller problems.
\lem
\label{lem:simple-non-sym}
\beqn
\label{first_step}
\mu^r_{[k]} \quad \le \quad \l \cdot \mu^{r-1}_{[k]} ~~+~~ (1-\l) \cdot \E_{i \in [k]}~ \mu^{r-1}_{[k] \setminus \{i\}}
\eeqn
\elem

\prf

\noi Since the feasible solution $\Big\{x^R_{S,i}\Big\}$ satisfies constraints 2 and 3 of (\ref{OP}), for any $i \in R \subseteq [k]$ holds $\mu^R\Big([k]\Big) ~=~ \mu^{R \setminus \{i\}}\Big([k] \setminus \{i\}\Big) ~+~ x^R_{[k],i}$.

\noi Similarly, $\mu^{R \setminus \{i\}}\Big([k]\Big) \quad = \quad \mu^{R \setminus \{i\}}\Big([k] \setminus \{i\}\Big) ~~+~~ x^{R \setminus \{i\}}_{[k],i}$.

\noi Hence, by constraint 4,
\[
x^R_{[k],i} ~\le~ \l \cdot \(x^{R \setminus \{i\}}_{[k],i}\) ~=~ \l \cdot \bigg(\mu^{R \setminus \{i\}}\Big([k]\Big) ~-~ \mu^{R \setminus \{i\}}\Big([k] \setminus \{i\}\Big)\bigg)
\]

\noi Averaging,
\[
\mu^r_{[k]} \quad = \quad \E_{R \subseteq [k], ~|R| = r} ~~\mu^R_{[k]} \quad = \quad \E_{R,~i \in R} ~~\bigg(\mu^{R \setminus \{i\}}\Big([k] \setminus \{i\}\Big) ~+~ x^R_{[k],i}\bigg) \quad \le
\]

\[
\E_{R,~i \in R} ~~\mu^{R \setminus \{i\}}\Big([k] \setminus \{i\}\Big) \quad + \quad \l \cdot \E_{R,~i \in R} ~~ \bigg(\mu^{R \setminus \{i\}}\Big([k]\Big) ~-~ \mu^{R \setminus \{i\}}\Big([k] \setminus \{i\}\Big)\bigg) \quad =
\]
\[
\l \cdot \E_{R,~i \in R} ~~ \mu^{R \setminus \{i\}}\Big([k]\Big) ~~+~~ (1-\l) \cdot \E_{R,~i \in R} ~~\mu^{R \setminus \{i\}}\Big([k] \setminus \{i\}\Big)
\]

\noi It remains to note
\[
\E_{R,~i \in R} ~~\mu^{R \setminus \{i\}}\Big([k] \setminus \{i\}\Big) \quad = \quad \E_{i \in [k]}~ \E_{|T| = r-1,~T \subseteq [k] \setminus \{i\}} ~~\mu^T\Big([k] \setminus \{i\}\Big) \quad = \quad \E_{i \in [k]}~ \mu^{r-1}_{[k] \setminus \{i\}}
\]
and, similarly, $\E_{R,~i \in R} ~\mu^{R \setminus \{i\}}\Big([k]\Big) ~=~ \mu^{r-1}_{[k]}$.

\eprf
\ignore{
\beqn
\label{first_step}
\mu^r_{[k]} \quad \le \quad \E_i~ \mu^{r-1}_{[k] \setminus \{i\}} ~~+~~ \phi\(\mu^{r-1}_{[k]} ~-~ \E_i~ \mu^{r-1}_{[k] \setminus \{i\}}\)
\eeqn
}

\noi We now prove (\ref{gen_ineq}), starting from (\ref{first_step}).

\noi First, note that, by Lemma~\ref{lem:sol-restrict} and by the induction hypothesis for $k-1$, we have $\mu^{r-1}_{[k] \setminus \{i\}} ~\le~ {\cal S}[\mu]^{r-1}_{[k] \setminus \{i\}}$, for all $i \in [k]$.

\noi Next, note that, by the induction hypothesis for $k$ and $r-1$, we have $\mu^{r-1}_{[k]}  ~\le~ {\cal S}[\mu]^{r-1}_{[k]}$.

\noi This gives
\[
\mu^r_{[k]} \quad \le \quad \l \cdot {\cal S}[\mu]^{r-1}_{[k]}  ~~+~~ (1-\l) \cdot \E_{i \in [k]}~ {\cal S}[\mu]^{r-1}_{[k] \setminus \{i\}}
\]

\noi This implies that to prove (\ref{gen_ineq}) it suffices to show the following two identities:

\begin{enumerate}
\item
\[
\E_{i \in [k]}~ {\cal S}[\mu]^{r-1}_{[k] \setminus \{i\}} \quad = \quad {\cal S}[\mu]^{r-1}_{[k]}\Big(k-1\Big)
\]

\item
\[
{\cal S}[\mu]^{r}_{[k]} \quad = \quad  \l \cdot {\cal S}[\mu]^{r-1}_{[k]} ~~+~~ (1-\l) \cdot {\cal S}[\mu]^{r-1}_{[k]}\Big(k-1\Big)
\]

\end{enumerate}

\lem
\label{lem:tech1}
\[
\E_{i \in [k]}~ {\cal S}[\mu]^{r-1}_{[k] \setminus \{i\}} \quad = \quad {\cal S}[\mu]^{r-1}_{[k]}\Big(k-1\Big)
\]
\elem
\prf
We introduce the following notation. For $i = 1,...,k$ and for $0 \le r < s \le k - 1$, let
\[
y_{s,i} \quad = \quad y_{s,~ [k] \setminus \{i\}} \quad \mbox{and} \quad x^r_{s,i} \quad = \quad x^r_{s,~ [k] \setminus \{i\}}
\]

\noi The values on the RHS of these identities are defined as in (\ref{y_S}) and in (\ref{x_S}) for the corresponding restricted problems.

\noi We start with observing that $\E_{i \in [k]} y_{s,i} ~=~ y_s$. In fact, by definition,
\[
\E_{i \in [k]} ~y_{s,i}\quad  =  \quad \E_{i \in [k]} ~\E_{|S| = s,S \subseteq [k] \setminus \{i\}, j \in S} ~~y_{S,j} \quad = \quad \E_{|S| = s, j \in S} ~~y_{S,j} \quad = \quad y_s
\]

\noi Next, we claim that for all $0 \le r < s \le k-1$ holds $\E_{i \in [k]} ~x^r_{s,i} ~=~  x^r_s$.

\noi This is easy to verify by induction on $r$.  Note that we already know the claim holds for $r = 0$, and the induction step follows directly from the definitions and the induction hypothesis.

\noi We now apply (\ref{value-sym}) to the restricted problems, to obtain that, for each $1 \le i \le k$ holds
\[
{\cal S}[\mu]^{r-1}_{[k] \setminus \{i\}} \quad = \quad \sum_{j=1}^{k - r} y_{j,~i} \quad + \quad \l \cdot \sum_{t=0}^{r-2} x^t_{k-r+t+1,~i}
\]
Hence, we have:
\[
\E_{i \in [k]}~ {\cal S}[\mu]^{r-1}_{[k] \setminus \{i\}} ~~=~~  \sum_{j=1}^{k - r} \E_{i \in [k]} ~y_{j,~i} ~~+~ \quad \l \cdot \sum_{t=0}^{r-2} \E_{i \in [k]} ~x^t_{k-r+t+1,~i} ~~=~~ \sum_{j=1}^{k - r} ~y_j ~+~ \l \cdot \sum_{t=0}^{r-2}  ~x^t_{k-r+t+1}
\]

\noi This, by (\ref{full-sym-k-1}), equals to ${\cal S}[\mu]^{r-1}_{[k]}(k-1)$, completing the proof of the lemma.

\eprf

\lem
\label{lem:tech2}
\[
{\cal S}[\mu]^{r}_{[k]} \quad = \quad  \l \cdot {\cal S}[\mu]^{r-1}_{[k]} ~~+~~ (1-\l) \cdot {\cal S}[\mu]^{r-1}_{[k]}\Big(k-1\Big)
\]
\elem
\prf

\noi The proof of this lemma is similar to that of Lemma~\ref{lem:simple-non-sym}.

\noi Since the symmetric solution ${\cal S}_{[k]}\Big[x^R_{S,i}\Big]$ \Big( which is the same as $\Big\{\bar{x}^R_{S,i}\Big\}$\Big) satisfies constraints 2 and 3 of (\ref{OP}), for any $i \in R \subseteq [k]$ holds
\[
{\cal S}[\mu]_{[k]}^R\Big([k]\Big) \quad = \quad {\cal S}[\mu]_{[k]}^{R \setminus \{i\}}\Big([k] \setminus \{i\}\Big) \quad + \quad  {\cal S}_{[k]}\Big[x^R_{[k],i}\Big]
\]

\noi Consider the notation we have introduced above. Using items 3 and 4 in the description of this notation, and recalling ${\cal S}_{[k]}\Big[x^R_{[k],i}\Big] = \l \cdot x^{r-1}_k$, we can rewrite this equality as
\[
{\cal S}[\mu]^{r}_{[k]} \quad = \quad {\cal S}[\mu]^{r-1}_{[k]}\Big(k-1\Big) ~~+~~ \l \cdot x^{r-1}_k
\]

\noi On the other hand, we have, for $i \in R \subseteq [k]$:
\[
{\cal S}[\mu]_{[k]}^{R \setminus \{i\}}\Big([k]\Big) \quad = \quad {\cal S}[\mu]_{[k]}^{R \setminus \{i\}}\Big([k] \setminus \{i\}\Big) \quad + \quad  {\cal S}_{[k]}\Big[x^{R \setminus \{i\}}_{[k],i}\Big]
\]
which is the same as
\[
{\cal S}[\mu]^{r-1}_{[k]}  \quad = \quad {\cal S}[\mu]^{r-1}_{[k]}\Big(k-1\Big) \quad + \quad x^{r-1}_k
\]
Combining these two identities immediately implies the claim of the lemma.

\eprf

\noi This completes the proof of (\ref{gen_ineq}) and of the theorem.

\eprf

\subsection{The value of the symmetric optimization problem}

\noi Let $\left\{\bar{x}^R_{S,i}\right\}$ be the symmetric solution for the symmetric version of (\ref{OP}). By Corollary~\ref{cor:symm-opt}, its value depends linearly on the symmetric boundary data $y_1,...,y_k$, since  $\{x^r_t\}$ are fixed linear functions of $y_1,...,y_k$. Let us denote this value by $V\(y_1,...,y_k\)$.

\noi For $1 \le s \le k$, let $e_s$ be the initial data vector with $y_s = 1$ and all the remaining $y_t$ vanishing. Then $V\(y_1,\ldots,y_k\) ~=~ \sum_{s=1}^k y_s \cdot V\(e_s\)$.

\noi Next, we find the values of the parameters $x^r_t$ for initial data given by a unit vector.

\lem
\label{lem:sym_parameters}
Let the initial data be given by the unit vector $e_s$, for some $1 \le s \le k$. Then the values of the parameters $x^r_t$, for $0 \le r < t \le k$, are as follows.

\[
x^r_t \quad = \quad \left\{\begin{array}{lll} {r \choose {t-s}} \cdot \l^{r - (t-s)} \cdot (1-\l)^{t-s} & \mbox{if} & s ~\le~ t ~\le~ s+r  \\
0 & \mbox{otherwise} & \end{array}\right.
\]

\noi (We use the convention ${0 \choose 0} = 1$.)
\elem
\prf
The claim of the lemma is easily verifiable by induction on $r$, or by directly verifying that (\ref{x_S}) holds.
\eprf

\cor
\label{cor:Vs}
\[
V\(e_s\) \quad = \quad \l^s \cdot \sum_{m=0}^{k-s} {{s+m-1} \choose m} \cdot (1-\l)^m \quad = \quad 1 ~~-~~ \sum_{j=0}^{s-1} {k \choose j} \l^j (1-\l)^{k-j}
\]
\ecor
\prf
The first equality follows from Corollary~\ref{cor:symm-opt}. For the second equality, we proceed as follows
\[
V\(e_s\) \quad = \quad \frac{\l^s}{(s-1)!} \cdot \frac{\partial^{s-1}}{\partial x^{s-1}} ~~\bigg[(1 + x + \ldots + x^{k-1}\bigg]_{x = 1-\l} \quad =
\]
\[
\frac{\l^s}{(s-1)!} \cdot \Bigg(\frac{\partial^{s-1}}{\partial x^{s-1}} ~~ \bigg[\frac{1}{1-x}\bigg]_{x = 1 - \l} ~~-~~ \frac{\partial^{s-1}}{\partial x^{s-1}} ~~\bigg[\frac{x^k}{1-x}\bigg]_{x = 1 - \l}\Bigg) \quad =
\]
\[
1 \quad - \quad \frac{\l^s}{(s-1)!} \cdot \frac{\partial^{s-1}}{\partial x^{s-1}} ~~\bigg[\frac{x^k}{1-x}\bigg]_{x = 1 - \l}
\]
We have
\[
\frac{\partial^{t}}{\partial x^t} ~~\bigg[\frac{x^k}{1-x}\bigg] \quad = \quad \sum_{i=0}^t {t \choose i} \frac{\partial^{i}}{\partial x^i} ~\bigg[\frac{1}{1-x}\bigg] \cdot \frac{\partial^{t-i}}{\partial x^{t-i}} ~\Big[x^k\Big] \quad =
\]
\[
\sum_{i=0}^t {t \choose i} \cdot i! \cdot \frac{k!}{(k-t+i)!} \cdot x^{k-t+i} \cdot \frac{1}{(1-x)^{i+1}}
\]
Substituting $j = t - i$ and rearranging, this is
\[
\frac{t!}{(1-x)^{t+1}} \cdot \sum_{j=0}^t {k \choose j} (1-x)^j \cdot  x^{k-j}
\]
Substituting $t = s - 1$, $x = 1-\l$, and simplifying, we get
\[
V\(e_s\) \quad = \quad 1 \quad - \quad \sum_{j=0}^{s-1} {k \choose j} \l^j (1-\l)^{k-j}
\]

\eprf

\cor
\label{cor:sym_val}
\[
V\(y_1,\ldots,y_k\) \quad = \quad \sum_{s=1}^k ~~\(1 ~-~  \sum_{j=0}^{s-1} {k \choose j} \l^j (1-\l)^{k-j}\) ~\cdot~ y_s
\]
\ecor

\section{Proof of Theorem~\ref{thm:main}}
\label{sec:proof-average}

\noi We start with introducing some more notation.

\subsubsection{Notation}
\begin{itemize}

\item
\noi For a subset $S$ of $[n]$ of cardinality at most $n-2$, and for distinct $i, j \not \in S$, we set
\[
Z_{S;i,j} ~~=~~ Ent\Big(f~|~S \cup \{i,j\}\Big) ~-~ Ent\Big(f~|~S \cup \{i\} \Big) ~-~ Ent\Big(f~|~S \cup \{j\} \Big) ~+~ Ent\Big(f~|~S\Big)
\]

\item

\noi For $s = 1,...,n-1$, let $t_s ~=~ \E_{S,i,j} ~Z_{S;i,j}$.

\noi Here the expectation is taken over all subsets $S$ of $[n]$ of cardinality $s-1$, and, given $S$, over all distinct $i,j$ not in $S$.

\item

\noi Let $A$ be a subset of $[n]$ of cardinality $k < n$ and let $m \not \in A$. For $1 \le s \le k$, let
\[
Y(A,m,s) \quad = \quad \E_{S,i} ~Z_{S;i,m}
\]
where the expectation goes over subsets $S \subseteq A$ of cardinality $s-1$, and over $i \in A \setminus S$.

\item

\noi For $1 \le s \le k \le n$ let
\[
\Lambda(k,s,\l) \quad = \quad 1 ~~-~~ \sum_{j=0}^{s-1} {k \choose j} \l^j (1-\l)^{k-j}
\]

\end{itemize}

\pro
\label{pro:mi-noisy}
Let $f$ be a nonnegative function on $\H$. Let $A$ be a subset of $[n]$ of cardinality $k < n$ and let $m \not \in A$.

\noi Then
\[
I_{T_{e_A} f}~(A,m) \quad \le \quad \sum_{s=1}^k ~~\Lambda(k,s,\l) ~\cdot~ Y(A,m,s)
\]
\epro

\prf

\noi By Theorem~\ref{thm:mi_opt}, the value of $I_{T_{e_A} f}~(A,m)$ is bounded by the value of the linear optimization problem (\ref{OP}), with appropriate changes of indices.

\noi By Theorem~\ref{thm:symm-opt}, this last value is upperbounded by the value of the symmetric version of the problem, which, according to Corollary~\ref{cor:sym_val}, and tracing out the appropriate changes in indices and notation, is given by $\sum_{s=1}^k ~\Lambda(k,s,\l) \cdot Y(A,m,s)$.

\eprf

\ignore{
\thm
\label{thm:main}
Let $f$ be a nonnegative function on the cube. Let $v = \l \cdot n + \sqrt{n \ln n}$.

\noi Then
\[
Ent\Big(T_{\e} f\Big) ~~\le~~ \sum_{i=1}^n \phi\bigg(Ent\Big(f~|~\{i\}\Big)\bigg) ~+~ \E_B ~Ent\Big(f~|~B\Big) ~-~ \frac{v}{n} \cdot \sum_{i=1}^n Ent\Big(f~|~\{i\}\Big) ~+
\]
\[
O\(\sqrt{\frac{\log n}{n}}\) \cdot Ent\Big(f\Big)
\]
Where the expectation in the second summand on the RHS is taken over subsets $B$ of $[n]$ of cardinality $\lceil v \rceil$.
\ethm
}

\prf (Of the theorem)

\noi The proof relies on several lemmas. We start with a technical claim.

\lem
\label{lem:w_s}
Let $1 \le s \le n-1$ be integer parameters. Let $0 < \l < 1$. Then
\[
\sum_{k=s}^{n-1}  \Lambda(k,s,\l) \quad = \quad \bigg(n - \frac{s}{\l}\bigg) ~+~ \frac{1}{\l} \cdot \sum_{j=0}^{s-1} ~\sum_{t=0}^j {n \choose t} \l^t (1-\l)^{n-t}
\]
\elem
\prf
\[
\sum_{k=s}^{n-1}  \Lambda(k,s,\l) \quad = \quad \sum_{k=s}^{n-1} \Bigg(1 ~~-~~ \sum_{j=0}^{s-1} {k \choose j} \l^j (1-\l)^{k-j}\Bigg) \quad =
\]
\[
\Big(n-s\Big) ~~-~~ \sum_{k=s}^{n-1} ~\sum_{j=0}^{s-1} {k \choose j} \l^j (1-\l)^{k-j} \quad = \quad \Big(n-s\Big) ~~-~~ \sum_{j=0}^{s-1} ~\l^j  \cdot \sum_{k=s}^{n-1} ~ {k \choose j} (1-\l)^{k-j}
\]

\noi A simple calculation, similar to that in the proof of Corollary~\ref{cor:Vs}, gives
\[
\l^j  \cdot \sum_{k=s}^{n-1} ~ {k \choose j} (1-\l)^{k-j} \quad = \quad \frac{1}{\l} \cdot \Bigg(\sum_{t=0}^j {s \choose t} \l^t (1-\l)^{s-t} ~~-~~\sum_{t=0}^j {n \choose t} \l^t (1-\l)^{n-t}\Bigg)
\]
The proof of the lemma is completed by summing the RHS over $j$, and observing
\[
\sum_{j=0}^{s-1} ~\sum_{t=0}^j {s \choose t} \l^t (1-\l)^{s-t} \quad = \quad (1-\l) \cdot s
\]

\eprf

\lem
\label{lem:noisy_ts}
Let $f$ be a nonnegative function on $\H$ with expectation $1$. Then
\[
Ent\Big(T_{\e} f\Big) \quad \le \quad \sum_{i=1}^n \phi\bigg(Ent\Big(f~|~\{i\}\Big)\bigg) ~~+~~\sum_{s=1}^{n-1} w_s \cdot t_{s}
\]
where
\[
w_s \quad = \quad \Big(\l n - s\Big) ~+~ \sum_{j=0}^{s-1} ~\sum_{t=0}^j {n \choose t} \l^t (1-\l)^{n-t}
\]
\elem

\lem
\label{lem:ts}
Let $f$ be a nonnegative function on $\H$. For any $0 \le u \le n-1$ holds
\[
\E_{|B| = u+1} ~Ent\Big(f~|~B\Big)  ~-~  (u+1) \cdot \E_{i \in [n]} ~Ent\Big(f~|~\{i\}\Big) \quad = \quad \sum_{s=1}^u \Big(u-s+1\Big) \cdot t_{s}
\]
\elem

\noi Next, we derive the theorem, assuming Lemmas~\ref{lem:noisy_ts} and~\ref{lem:ts} to hold.

\noi Let $T$ be a random subset of $[n]$ generated by sampling each element $i \in [n]$ independently with probability $\l$. We will show
\[
\E_T ~\bigg(Ent\Big(f~|~T\Big) ~-~ \sum_{i \in T} Ent\Big(f~|~\{i\}\Big)\bigg) \quad = \quad \sum_{s=1}^{n-1} w_s \cdot t_s
\]

\noi Combining this with the claim of Lemma~\ref{lem:noisy_ts} will complete the proof.

\noi For $0 \le k \le n$, let $p_k = {n \choose k} \l^k (1-\l)^{n-k}$. And, for $0 \le u \le n - 1$, let
\[
\mu_u ~~=~~ \E_{|B| = u+1} ~\bigg(Ent\Big(f~|~B\Big) ~-~ \sum_{i \in B} ~Ent\Big(f~|~\{i\}\Big) \bigg)
\]

\noi Then, using Lemma~\ref{lem:ts} and observing that $\mu_0 = 0$,
\[
\E_T ~\bigg(Ent\Big(f~|~T\Big) - \sum_{i \in T} Ent\Big(f~|~\{i\}\Big)\bigg) ~=~ \sum_{k=2}^n p_k ~\mu_{k-1} ~=
\]
\[
\sum_{k = 2}^n p_k  ~\sum_{s=1}^{k-1} \(k-s\) \cdot t_s ~=~
\sum_{s=1}^{n-1} \(\sum_{k=s+1}^n \(k-s\) p_k  \) \cdot t_s
\]

\noi We conclude by verifying the identity $w_s ~=~ \sum_{k=s+1}^n \(k-s\) p_k$, for $s = 1,...,n-1$.

\noi In fact,
\[
w_s ~~=~~ \Big(\l n - s\Big) ~+~ \sum_{j=0}^{s-1} \sum_{t=0}^j ~p_t ~~=~~ \sum_{k=0}^n \(k-s\) p_k ~+~ \sum_{t=0}^{s-1} \(s-t\) p_t ~~=~~ \sum_{k=s+1}^n \(k-s\) p_k
\]

\eprf

\ignore{
\noi We are going to apply the Chernoff bound in the following form \cite{AS}:

\noi {\it Let $X_k \sim B(k,\l)$ be a Bernoulli random variable. Then for any $a \ge 0$ holds $Pr\Big\{\Big | X_k - \l k \Big | > a\Big\} \le e^{-2a^2/k}$.}

\noi Let us set $d_s = \sum_{j=0}^{s-1} ~\sum_{t=0}^j {n \choose t} \l^t (1-\l)^{n-t}$. Note that $d_s ~=~ \sum_{j=0}^{s-1} Pr\Big\{X_n \le j\Big\}$, and that $w_s = \Big(\l n - s\Big) + d_s$.

\rem
\label{rem:prob}
We note, for future use, the following two probabilistic interpretations for $w_s$:
\[
w_s ~~=~~ \Big(\l n - s\Big) ~+~  \sum_{j=0}^{s-1} Pr\Big\{X_n \le j\Big\} ~~\mbox{and}~~ w_s ~~=~~ \l \cdot \sum_{k=s}^{n-1}  \Lambda(k,s,\l) ~~=~~ \l \cdot \sum_{k=s}^{n-1} ~Pr\Big\{X_k \ge s\Big\}
\]
\erem

\noi Using the Chernoff bound for $X_n$ gives that for $s < \l n - \sqrt{n \ln n}$ holds
\[
d_s \quad = \quad \sum_{j=0}^{s-1} Pr\Big\{X_n \le j\Big\} \quad \le \quad \frac{1}{n}
\]

\noi Applying the bound to $X_k \sim B(k,\l)$ gives that for $s > \l n + \sqrt{n \ln n}$ holds
\[
w_s \quad = \quad \l \cdot \sum_{k=s}^{n-1}  \Lambda(k,s,\l) \quad \le \quad \sum_{k=s}^{n-1} ~Pr\Big\{X_k > s\Big\} \quad < \quad \frac{1}{n}
\]

\noi In the first step we used Lemma~\ref{lem:w_s}.

\noi Finally, applying the bound to $X_n$ again, we have for $\l n - \sqrt{n \ln n} \le s \le \l n + \sqrt{n \ln n}$
\[
d_s \quad = \quad \sum_{j=0}^{s-1} Pr\Big\{X_n \le j\Big\} \quad = \quad \sum_{j=0}^{\l n - \sqrt{n \ln n}} Pr\Big\{X_n \le j\Big\} ~~+~ \sum_{j=\l n - \sqrt{n \ln n}+1}^{s-1} Pr\Big\{X_n \le j\Big\} \quad \le
\]
\[
O\Big(\sqrt{n \ln n}\Big)
\]

\noi Combining the two lemmas, taking $u = \l n + \sqrt{n \ln n} - 1$ in Lemma~\ref{lem:ts}, and setting $v = u + 1$, we have
\[
Ent\Big(T_{\e} f\Big) \quad \le \quad  \E_{|B| = v} ~Ent\Big(f~|~B\Big)  ~~-~~  \frac{v}{n} \cdot \sum_{i=1}^n ~Ent\Big(f~|~\{i\}\Big) ~~+~~ \sum_{i=1}^n \phi\bigg(Ent\Big(f~ \Big |~\{i\}\Big)\bigg) ~~+
\]
\[
\sum_{s=1}^{v} ~d_s \cdot t_{s}  ~~+~~ \sum_{s=v+1}^{n-1} w_s \cdot t_{s} \quad =
\]
\[
\E_{|B| = v} ~\bigg(Ent\Big(f~|~B\Big) ~-~ \sum_{i \in B} Ent\Big(f~|~\{i\}\Big) \bigg) ~~+~~ \sum_{i=1}^n \phi\bigg(Ent\Big(f~ \Big |~\{i\}\Big)\bigg) ~~+~~ E(n),
\]
where $E(n)$ is an error term.

\noi We now estimate $E(n)$. Observe that Lemma~\ref{lem:ts} applied with $u = n-1$ implies
\[
\sum_{s=1}^{n-1} (n-s) \cdot t_{s} \quad \le \quad Ent\Big(f\Big)
\]
In particular
\[
\sum_{s = \l n - \sqrt{n \ln n}}^{\l n + \sqrt{n \ln n}} ~~t_{s} \quad \le \quad O\(\frac 1n\) \cdot Ent\Big(f\Big)
\]

\noi Assuming $n \ge 10 \cdot \(\frac{1}{(1-\l)^2} \cdot \log\(\frac{1}{1-\l}\)\)$,  the asymptotic notation in the above expression hides an absolute constant.

\noi Hence, by the above discussion,
\[
\sum_{s=1}^{v} ~d_s \cdot t_{s} ~~=~~ \sum_{s=1}^{\l n - \sqrt{n \ln n}} ~~d_s \cdot t_{s} ~~+~ \sum_{s=\l n - \sqrt{n \ln n} +1}^{v} ~~d_s \cdot t_{s} ~~ \le ~~ O\(\sqrt{\frac{\ln n}{n}}\) \cdot Ent\Big(f\Big)
\]
where both summands in the middle expression are upperbounded by the maximum of $d_s$ on the corresponding domain, times the sum of $t_s$ on that domain.

\noi Similarly,
\[
\sum_{s=v+1}^{n-1} w_s \cdot t_{s} \quad \le \quad \max_{v+1 \le s \le n} ~w_s \cdot \sum_{s=v+1}^{n-1} t_{s} \quad \le \quad O\(\frac 1n\) \cdot Ent\Big(f\Big)
\]
Hence $E(n) ~\le~ O\(\sqrt{\frac{\ln n}{n}}\) \cdot Ent\Big(f\Big)$.

\noi This completes the proof of the theorem, given the lemmas hold.
}

\noi It remains to prove the lemmas.

\prf (Of Lemma~\ref{lem:noisy_ts})

\noi Recall that, by the chain rule for noisy entropy (\ref{chain-rule}), for any permutation $\sigma \in S_n$ holds that $Ent\Big(T_{\e} f\Big)$ is bounded from above by
\[
\sum_{i=1}^n \phi\Bigg( Ent\Big(T_{\e_{\left\{\sigma(1),\ldots,\sigma(i-1)\right\}}} f~|~\left\{\sigma(1),\ldots,\sigma(i)\right\}\Big) ~-~ Ent\Big(T_{\e_{\left\{\sigma(1),\ldots,\sigma(i-1)\right\}}} f~|~\left\{\sigma(1),\ldots,\sigma(i-1)\right\}\Big) \Bigg)
\]
Using the notation introduced in Subsection~\ref{subsubsec:improve}, we can write this as
\[
\sum_{i=1}^n \phi\Bigg(Ent\Big(f~|~\{\sigma(i)\}\Big) ~~+~~ I_{T_{\e_{\left\{\sigma(1),\ldots,\sigma(i-1)\right\}}} f}~\bigg(\Big\{\sigma(1),\ldots,\sigma(i-1)\Big\},~\sigma(i)\bigg)\Bigg)
\]
Observe that the function $\phi$ is concave, and $\phi(0) = 0$. Hence $\phi(x+y) \le \phi(x) + \phi(y)$ for any $0 \le x,y \le 1$. By this subbaditivity of $\phi$, the last expression is at most
\[
\sum_{i=1}^n \phi\bigg(Ent\Big(f~|~\{i\}\Big)\bigg) ~~+~~ \sum_{k=2}^n \phi\Bigg(I_{T_{\e_{\left\{\sigma(1),\ldots,\sigma(k-1)\right\}}} f}~\bigg(\Big\{\sigma(1),\ldots,\sigma(k-1)\Big\},~\sigma(k)\bigg)\Bigg)
\]
Averaging this expression over all $\sigma \in S_n$, we obtain
\[
Ent\Big(T_{\e} f\Big) \quad \le \quad \sum_{i=1}^n \phi\bigg(Ent\Big(f~|~\{i\}\Big)\bigg) ~~+~~ \mu,
\]
where
\[
\mu = \E_{\sigma} ~~\sum_{k=2}^n \phi\Bigg(I_{T_{\e_{\left\{\sigma(1),\ldots,\sigma(k-1)\right\}}} f}~\bigg(\Big\{\sigma(1),\ldots,\sigma(k-1)\Big\},~\sigma(k)\bigg)\Bigg)
\]

\noi Next, we upper bound $\mu$. By transitivity of action of the symmetric group and by concavity of $\phi$ we have
\[
\mu \quad \le \quad \sum_{k=1}^{n-1} \phi\Big(b_k\Big) \quad \mbox{where} \quad b_k \quad = \quad \E_{A, m} ~T_{\e_A f}~\Big(A,m\Big)
\]
where the expectation is over all $A \subseteq [n]$ of cardinality $k$ and $m \not \in A$.

\noi Applying Proposition~\ref{pro:mi-noisy}, we get
\[
b_k \quad \le \quad \E_{A,m} ~\sum_{s=1}^k \Lambda\Big(k,s,\l\Big) \cdot Y\Big(A,m,s\Big) \quad = \quad \sum_{s=1}^k \Lambda\Big(k,s,\l\Big) \cdot \E_{A,m} ~Y\Big(A,m,s\Big)
\]

\noi By the definition of $Y\Big(A,m,s\Big)$,
\[
\E_{A,m} ~Y\Big(A,m,s\Big) \quad = \quad \E_{A,m} ~\E_{S,i} ~Z_{S;i,m} \quad = \quad \E_{S,i,m} ~Z_{S;i,m} \cdot \E_A ~1 \quad = \quad \E_{S,i,m} ~Z_{S;i,m}
\]
where in the second expression the first expectation is over $k$-subsets $A$ of $[n]$ and $m \not \in A$, and the second expectation is over $(s-1)$-subsets $S$ of $A$ and over $i \in A \setminus S$. Rearranging, we get the third expression in which the first expectation is over all subsets $S$ of $[n]$ of cardinality $s-1$ and over all distinct $i, m \not \in S$, and the second expectation is over all supersets $A$ of $S$ of cardinality $k$ with $i \in A$ and $m \not \in A$.

\noi Recalling the definition of $t_s$ above, we deduce $b_k \quad = \quad \sum_{s=1}^k \Lambda\Big(k,s,\l\Big) \cdot t_{s}$.

\noi Using the inequality $\phi(x) \le \l x$, and Lemma~\ref{lem:w_s}, we have
\[
\mu \quad \le \quad \l \cdot \sum_{k=1}^{n-1} b_k \quad = \quad \l \cdot \sum_{k=1}^{n-1} ~\sum_{s=1}^k \Lambda\Big(k,s,\l\Big) \cdot t_{s} \quad =
\]
\[
\sum_{s=1}^{n-1} t_{s} \cdot \(\l \cdot \sum_{k=s}^{n-1} \Lambda\Big(k,s,\l\Big)\) \quad = \quad \sum_{s=1}^{n-1} w_s \cdot t_{s}
\]

\eprf

\prf (of Lemma~\ref{lem:ts})

\noi By (\ref{mi-chain}), for any subset $A$ of $[n]$ of cardinality $1 \le k \le n-1$, for any $m \not \in A$, and for any bijection $\tau: [k] \rarrow A$ holds, in the notation of this section,
\[
\sum_{s=1}^k ~Z_{\{\tau(1),\ldots,\tau(s-1)\};\tau(s),m} \quad = \quad I_f\Big(A,m\Big)
\]

\noi We now average over all the variables, setting
\[
c_k \quad = \quad \E_{A,m,\tau} ~\sum_{s=1}^k ~Z_{\{\tau(1),\ldots,\tau(s-1)\};\tau(s),m}
\]

\noi On one hand, we have
\[
c_k \quad = \quad \E_{A,m} ~I_f\Big(A,m\Big) \quad = \quad \E_{A,m} ~ \bigg(Ent\Big(f~|~A \cup \{m\}\Big) - Ent\Big(f~|~A\Big) - Ent\Big(f~|~\{m\}\Big)\bigg) \quad =
\]
\[
\E_{|B| = k+1} ~Ent\Big(f~|~B\Big) ~-~ \E_{|A| = k} ~Ent\Big(f~|~A\Big) ~-~  \E_{i \in [n]} ~Ent\Big(f~|~\{i\}\Big)
\]

\noi On the other hand, similarly to the computation in the preceding lemma, we have
\[
c_k \quad = \quad \sum_{s=1}^k ~\E_{A,m,\tau} ~Z_{\{\tau(1),\ldots,\tau(s-1)\};\tau(s),m} \quad = \quad \sum_{s=1}^k ~ \E_{A,S,i,m} ~Z_{S;i,m} \quad = \quad \sum_{s=1}^k t_{s}
\]
where the expectation in the third expression is over $k$-subsets $A$ of $[n]$, over $(s-1)$-subsets $S$ of $A$, over $m \not \in A$ and $i \in A \setminus S$.

\noi Hence, for any $1 \le u \le n-1$ holds
\[
\E_{|B| = u+1} ~Ent\Big(f~|~B\Big)  ~-~  (u+1) \cdot \E_{i \in [n]} ~Ent\Big(f~|~\{i\}\Big) \quad = \quad \sum_{k=1}^u c_k \quad = \quad \sum_{s=1}^u \Big(u-s+1\Big) \cdot t_{s}
\]
completing the proof of the lemma and of the theorem.

\eprf

\section{Proof of Theorem~\ref{thm:KC-ent-sn}}
\label{sec:kc}

\noi Let $\delta$ be the constant in the theorem. We will assume in the following argument that $\delta$ is sufficiently small.

\noi Let $0 < \e < 1/2$ be a noise parameter, such that $(1-2\e)^2 \le \delta$. Let $\l = (1-2\e)^2$.

\noi Let $f:~\H \rarrow \{0,1\}$ be a boolean function, satisfying the constraints of the theorem. Let $1 \le k \le n$ be the coordinate such that $|\widehat{f}(k)|$ is large. W.l.o.g. assume that $k = 1$ and that $\widehat{f}(1)$ is positive.

\noi We introduce some additional notation.

\subsubsection*{Notation}

\begin{itemize}

\item
Let $0 \le \alpha \le \delta$ be such that $\widehat{f}(1) = (1-\alpha) \cdot \E f$.

\item
Let $0 \le \beta \le \delta$ be such that $\E f = 1/2 -\beta$. Let $\gamma = \alpha + \beta$.

\item
If $\alpha \le \l$, we define $\t = \(\frac{1 - \l}{1 - \alpha}\)^2$, and define auxiliary noise $\e_{\t}$, such that $\Big(1 - 2\e_{\t}\Big)^2 = \tau$.
If $\alpha > \l$, we set $\tau = 1$ and $\e_{\t} = 0$.

\item
Let $\e_1$ be such that $T_{\e} = T_{\e_1} T_{\e_{\t}}$. Let $\l_1 = (1-2\e_1)^2$. Note that $\l = \tau \cdot \l_1$.

\item
 Let $h = T_{\e_{\t}} f$. Note that $T_{\e} f = T_{\e_1} h$, and hence $Ent\Big(T_{\e} f\Big) =  Ent\Big(T_{\e_1} h\Big)$.

\end{itemize}

\subsection{Proof of the first claim of the theorem}

\noi We start with applying Theorem~\ref{thm:main} to the function $h$ with noise $\e_1$. The theorem is stated for functions with expectation $1$. We modify it, using the linearity of entropy, to obtain
\[
Ent\Big(T_{\e_1} h\Big) ~~\le~~   \E_T ~\bigg(Ent\Big(h~|~T\Big) - \sum_{i \in T} Ent\Big(h~|~\{i\}\Big) \bigg) ~+~
\E h \cdot \sum_{i=1}^n \phi\Bigg(Ent\(\frac{h}{\E h}~ \Big |~\{i\}\),~\e_1\Bigg)
\]

\noi Here $T$ is a random subset of $[n]$ generated by sampling each element $i \in [n]$ independently with probability $\(1-2\e_1\)^2$.

\noi Since there are more than one noise parameters involved, we now write the function $\phi$ with the noise parameter stated explicitly.

\noi Next, note that by (\ref{phi-lambda}), for any $1 \le i \le n$ holds
\[
\E h \cdot \phi\Bigg(Ent\(\frac{h}{\E h}~ \Big |~\{i\}\),~\e_1\Bigg) ~~\le~~ \l_1 \cdot Ent\Big(h~|~\{i\}\Big)
\]
Hence the previous inequality implies
\[
Ent\Big(T_{\e_1} h\Big) ~\le~  \l_1 \cdot \E_{T,1 \in T} \bigg(Ent\Big(h~|~T\Big) - Ent\Big(h~|~\{1\}\Big) \bigg)  ~~+
\]
\beqn
\label{stab}
\(1 - \l_1\) \cdot \E_{T,1 \not \in T} ~Ent\Big(h~|~T\Big) ~~+~~\E h \cdot \phi\Bigg(Ent\(\frac{h}{\E h}~ \Big |~\{1\}\),~\e_1\Bigg)
\eeqn

\noi The proof will be based on three lemmas, which upperbound each of the three summands on the RHS of (\ref{stab}).

\lem
\label{lem:1-in-B}
\[
\E_{T,1 \in T} ~\bigg(Ent\Big(h~|~T\Big) - Ent\Big(h~|~\{1\}\Big) \bigg) \quad \le \quad O\Bigg(\l_1 \cdot \gamma ~+~ \gamma^2 \ln\(\frac{1}{\gamma}\)\Bigg)
\]
\elem

\lem
\label{lem:1-not-in-B}
\[
\E_{T,1 \not \in T} ~Ent\Big(h~|~T\Big) \quad \le \quad O\Bigg(\l^2_1 \cdot \gamma ~+~ \l_1 \cdot \gamma^2 \ln\(\frac{1}{\gamma}\)\Bigg)
\]
\elem

\lem
\label{lem:marginal-1}
\[
\E h \cdot \phi\Bigg(Ent\(\frac{h}{\E h}~ \Big |~\{1\}\),~\e_1\Bigg) \quad \le \quad  \frac12 \cdot \Big(1 - H(\e)\Big) ~-~ \Omega\Big(\l \cdot \gamma \Big)
\]

\elem

\noi The asymptotic notation in each of the lemmas hides absolute constants.

\noi Given the lemmas, the first claim of the theorem is easy to verify. Indeed, recall that $\l_1$ is a constant multiple of $\l$. Hence, the lemmas and (\ref{stab}) imply that
\[
Ent\Big(T_{\e} f\Big) ~~=~~ Ent\Big(T_{\e_1} h\Big) ~~\le~~ \frac12 \cdot \Big(1 - H(\e)\Big)  ~-~ \Omega\Big(\l \cdot \gamma \Big) ~+~ o_{\l, \gamma \rarrow 0} \Big(\l \cdot \gamma \Big)
\]

\noi Therefore, for a sufficiently small $\delta > 0$, bearing in mind that $0 \le \alpha, \beta, \l \le \delta$, the claim holds.

\noi It remains to prove the lemmas. For that purpose we will need the following version of the logarithmic Sobolev inequality for the boolean cube.
\lem
\label{lem:LS-1}
Let $g$ be a nonnegative function on $\H$. Let ${\cal E}(g,g)$ be the {\it Dirichlet form}, given by ${\cal E}(g,g) = \E_{x \in \H} \E_{y \sim x} \Big(g(y) - g(x)\Big)^2$. Then
\[
{\cal E}(g,g) \quad \ge \quad 2 \ln 2 \cdot \E g \cdot Ent\Big(g\Big)
\]
\elem
\prf

\noi We start with a simple auxiliary claim.

\noi Let $x_1 \ge x_2 \ge ... \ge x_N$ be nonnegative numbers summing to $1$. Then the numbers $y_k = \frac{x^2_k}{\sum_{i=1}^N x^2_i}$, for $k = 1,...,N$, majorize $\{x_k\}$. That is,
\[
y_1 ~\ge~x_1,~~y_1 + y_2 ~\ge~x_1 + x_2,~~\ldots~~, y_1 + ... + y_N ~=~ 1 ~=~ x_1 + ... + x_N
\]

\noi To see this, fix some $1 \le t \le N$. We have to show $\sum_{k=1}^t x^2_k ~\ge~ \(\sum_{k=1}^t x_k \) \cdot \(\sum_{k=1}^N x^2_k\)$.

\noi We may and will assume that all of the $x_k$ are strictly positive. After some rearrangement, the claim reduces to showing
\[
\frac{\sum_{k=1}^t x^2_k}{\sum_{k=1}^t x_k} \quad \ge \quad \frac{\sum_{m=t+1}^N x^2_m}{\sum_{k=t+1}^N x_m}
\]
This holds because the LHS is lowerbounded by $x_t$, and the RHS is upperbounded by $x_{t+1}$.

\noi A simple corollary of this claim is that for any nonnegative not identically zero function $g$ on a finite domain endowed with uniform measure, holds that $g^2/\E g^2$ majorizes $g/ \E g$.

\noi This is well-known to imply (see \cite{MO}) that $g/ \E g$ is a convex combination of permuted versions of $g^2/\E g^2$. Since the entropy functional is linear and convex, this implies
\[
Ent\Big(g^2\Big) \quad \ge \quad \frac{\E g^2}{\E g} \cdot Ent\Big(g\Big) \quad \ge \quad \E g \cdot Ent\Big(g\Big)
\]

\noi The claim of the lemma follows from this inequality combined with the logarithmic Sobolev inequality \cite{Ledoux}:
\[
{\cal E}(g,g) \quad \ge \quad 2 \ln 2 \cdot Ent\Big(g^2\Big)
\]
\eprf

\noi In the following argument we are going to use the Walsh-Fourier expansion for functions on the boolean cube, writing a function $g$ as $\sum_{S \subseteq [n]} \widehat{g}(S) \cdot W_S$, where $\Big\{W_S\Big\}_{S \subseteq [n]}$ is the Walsh-Fourier basis.

\noi In particular, for the Dirichlet form, we have ${\cal E}(g,g) = 4 \cdot \sum_{S \subseteq [n]} |S| \widehat{g}^2(S)$. Hence the preceding lemma implies
\beqn
\label{LS-F}
Ent\Big(g\Big) \quad \le \quad \frac{2}{\ln 2} \cdot \frac{1}{\E g} \cdot \sum_{S \subseteq [n]} |S| ~\widehat{g}^2(S)
\eeqn

\noi We will also need the following precise version of an inequality of \cite{FKN}, due to \cite{JOW}:

\thm
\label{thm:JOW}
There exists a universal constant $L > 0$ with the following property. For $g:~\H \rarrow \{-1,1\}$, let $\rho = \(\sum_{A \subseteq [n]:|A| \ge 2} \widehat{g}^2(A)\)^{1/2}$. Then there exists some $B \subseteq [n]$ with $|B| \le 1$ such that
\[
\sum_{A \subseteq [n]:|A| \le 1,A \not = B} \widehat{g}^2(A) \quad \le \quad L \cdot \rho^4 \ln \(\frac{2}{\rho}\)
\]
and $|\widehat{g}(B)|^2 \ge 1 ~-~ \rho^2 ~-~ L \cdot \rho^4 \ln \(\frac{2}{\rho}\)$.

\ethm

\noi Consider the function $f$ and recall that it satisfies the assumptions of Theorem~\ref{thm:KC-ent-sn}.

\noi Let $g = 2 f - 1$. Then $g:~\H \rarrow \{-1,1\}$. Note that $\widehat{g}(0) = 2 \widehat{f}(0) - 1$, and that $\widehat{g}(S) = 2 \widehat{f}(S)$, for $|S| > 0$.

\noi In particular, $ \widehat{g}(0) =  2 \E f - 1 = -2 \beta$, and $\widehat{g}(\{1\}) = 2 (1 - \alpha) \E f = (1 - \alpha) (1 - 2 \beta)$.

\noi Recall that $0 \le \alpha, \beta \le \delta$, and that $\gamma = \alpha + \beta$. Hence, assuming $\delta$ is sufficiently small, we have
\beqn
\label{fl>2}
\sum_{|A| \ge 2} \widehat{f}^2(A) ~~\le~~ \sum_{|A| \ge 2} \widehat{g}^2(A) ~~\le~~ 1 ~-~ \widehat{g}^2(\{1\}) ~~\le~~ L \cdot \gamma,
\eeqn
for some absolute constant $L$.

\noi Applying Theorem~\ref{thm:JOW} to the function $g$, we get, for a sufficiently large constant $L_1$,
\beqn
\label{fl+others}
\sum_{k=2}^n \widehat{f}^2\Big(\{k\}\Big) \quad \le \quad  \sum_{k=2}^n \widehat{g}^2\Big(\{k\}\Big) \quad \le \quad L_1 \cdot \gamma^2 \ln\(\frac{1}{\gamma}\)
\eeqn

\subsubsection*{Proof of Lemma~\ref{lem:1-not-in-B}}

\noi Fix $T \subseteq [n]$. Let $g_T = \E\Big(h~|~T\Big)$.

\noi Note that $g_T = \sum_{S \subseteq T} \widehat{h}(S) \cdot W_S$, and hence, by (\ref{LS-F}), we have
\[
Ent\Big(g_T\Big) \quad \le \quad \frac{2}{\ln 2} \cdot \frac{1}{\E g_T} \cdot \sum_{S \subseteq T} |S| ~\widehat{h}^2(S) \quad = \quad \frac{2}{\ln 2} \cdot \frac{1}{\E h} \cdot \sum_{S \subseteq T} |S| ~\widehat{h}^2(S)
\]
Hence,
\[
\E_{T,1 \not \in T} ~Ent\Big(g_T\Big) ~~\le~~  \frac{2}{\ln 2} \cdot \frac{1}{\E h} \cdot ~\E_{T,1 \not \in T} \sum_{S \subseteq T} |S| ~\widehat{h}^2(S) ~~=~~
\frac{2}{\ln 2} \cdot \frac{1}{\E h} \cdot \sum_{S, 1 \not \in S} |S| \l_1^{|S|} ~\widehat{h}^2(S)
\]

\noi Recall that $h = T_{\e_\t} f$. This means (see, e.g., \cite{OD}) that for any $S \subseteq [n]$, holds $\widehat{h}(S) = \tau^{|S|/2} \cdot \widehat{f}(S)$.

\noi In particular, $|\widehat{h}(S)| \le |\widehat{f}(S)|$. Applying (\ref{fl>2}) and (\ref{fl+others}), we have that, for a sufficiently large absolute constant $L$, the last expression is bounded by
\[
L \cdot \bigg( \l^2_1 \cdot \gamma ~+~ \l_1 \cdot \gamma^2 \ln \(\frac{1}{\gamma}\) \bigg)
\]

\noi This concludes the proof of the lemma.
\eprf

\subsubsection*{Proof of Lemma~\ref{lem:marginal-1}}

\noi Let $g = \E\(\frac{f}{\E f}~\Big | ~\{1\}\)$. Then $g$ is a function on a $2$-point space $\{0,1\}$, with $g(0) = 2 - \alpha$ and $g(1) = \alpha$.

\noi Observe that the noise operator commutes with the projection operator. Hence, since $h = T_{\e_{\t}} f$, we have $g_1 := \E\(\frac{h}{\E h}~\Big | ~\{1\}\) = T_{\e_{\t}} g$.

\noi Observe also that, by the definition of Mrs. Gerber's function $\phi$, we have
\[
\phi\Bigg(Ent\(\frac{h}{\E h}~ \Big |~\{1\}\),~\e_1\Bigg) ~~=~~  Ent\Big(T_{\e_1} g_1 \Big) ~~=~~ Ent\Big(T_{\e_1} T_{\e_{\t}} g \Big) ~~=~~
Ent\Big(T_{\e} g \Big)
\]

\noi The last equality follows from the definition of $\e_1$ and $\e_{\t}$.

\noi It is easy to verify that $T_{\e} g(0) = 1 + (1-\alpha) \cdot \l^{1/2}$ and that $T_{\e} g(1) = 1 - (1-\alpha) \cdot \l^{1/2}$.

\noi Hence, $Ent\Big(T_{\e} g \Big) = 1 - H_2\(\frac{1 - (1-\alpha) \cdot \l^{1/2}}{2}\)$.

\noi Recall that
\[
H_2\(\frac{1-x}{2}\) \quad = \quad 1 ~~-~~ \frac{1}{\ln 2} \cdot \sum_{k=1}^{\infty} \frac{1}{2k(2k-1)} \cdot x^{2k}
\]
with the series converging absolutely for $-1 \le x \le 1$.

\noi Let $F(x) = 1 - H_2\(\frac{1-\sqrt{x}}{2}\)$, for $0 \le x \le 1$. Then $F(x) = \frac{1}{\ln 2} \cdot \sum_{k=1}^{\infty} \frac{1}{2k(2k-1)} \cdot x^{k}$.

\noi This is a convex function on $[0,1]$, and hence for any $0 \le x < y \le 1$ holds $F(y) - F(x) \ge (y - x) \cdot F'(x)$. The derivative $F'$ is given by $F'(x) = \frac{1}{2 \ln 2} \cdot \sum_{k=1}^{\infty} \frac{1}{2k-1} \cdot x^{k-1}$, with the series converging for $0 \le x < 1$.

\noi Hence $F' \ge \frac{1}{2 \ln 2}$ on $(0, 1)$, and $F(y) - F(x) \ge \frac{1}{2 \ln 2} \cdot (y - x)$. Applying this with $y = \l$ and $x = (1-\alpha)^2 \cdot \l$, we get
\[
\Big(1 - H_2(\e)\Big) ~-~ Ent\Big(T_{\e} g \Big) \quad = \quad F(\l) ~-~ F\Big((1-\alpha)^2 \cdot \l\Big) \quad \ge \quad c_1 \cdot \l \cdot \alpha
\]
where $c_1 > 0$ is an absolute constant.

\noi In other words,
\[
\phi\Bigg(Ent\(\frac{h}{\E h}~ \Big |~\{1\}\),~\e_1\Bigg) ~~=~~ Ent\Big(T_{\e} g \Big) ~~\le~~ \Big(1 - H_2(\e)\Big) ~-~ c_1 \cdot \l \cdot \alpha
\]

\noi To conclude the proof of the lemma, note that, for a sufficiently small $\l$, we have $Ent\Big(T_{\e} g \Big) \ge c_2 \cdot \l$, for an absolute constant $c_2$, and hence
\[
\E h \cdot \phi\Bigg(Ent\(\frac{h}{\E h}~ \Big |~\{1\}\),~\e_1\Bigg) ~~=~~ \(\frac12 - \beta\) \cdot Ent\Big(T_{\e} g \Big) ~~\le~~
\]
\[
\frac12 \cdot \Big(1 - H_2(\e)\Big)  ~-~  c \cdot \l \cdot (\alpha + \beta) ~~=~~ \frac12 \cdot \Big(1 - H_2(\e)\Big)  ~-~  c \cdot \l \cdot \gamma
\]
for an absolute constant $c$. For the inequality, note that $1 - H_2(\e) = F(\l) \ge \frac{1}{2\ln2} \cdot \l$.

\noi This completes the proof of the lemma. \eprf

\noi The proof of Lemma~\ref{lem:1-in-B} is somewhat harder. We present it in the next subsection.

\subsubsection{Proof of Lemma~\ref{lem:1-in-B}}

\noi We proceed similarly to the proof of Lemma~\ref{lem:1-not-in-B}, and use the notation introduced in that proof.

\noi Given a function $g$ on the boolean cube, we write $\E \Big(g~|~x_1 = 0,x_2,...,x_k\Big)$ for the restriction of $\E \Big(g~|~x_1,x_2,...,x_k\Big)$ on the subcube $x_1 = 0$, and similarly for $\E \Big(g~|~x_1 = 1,x_2,...,x_k\Big)$.

\noi We note that for $g = \sum_{S \subseteq [n]} \widehat{g}(S) \cdot W_S$, we have
\[
\E \Big(g~|~x_1 = 0,x_2,...,x_n\Big) \quad = \quad \sum_{R \subseteq [n], 1 \not \in R} ~\Big(\widehat{g}(R) ~+~ \widehat{g}(R \cup \{1\}) \Big) \cdot W_R
\]
and
\[
\E \Big(g~|~x_1 = 1,x_2,...,x_n\Big) \quad = \quad \sum_{R \subseteq [n], 1 \not \in R} ~\Big(\widehat{g}(R) ~-~ \widehat{g}(R \cup \{1\}) \Big) \cdot W_R
\]

\noi We will also use the following easily verifiable identity, holding for nonnegative functions $g$:
\[
Ent\Big(g\Big) ~-~ Ent\Big(g~|~\{1\}~\Big) \quad = \quad \frac12 \cdot Ent\Big(g~|~x_1 = 0,x_2,...,x_n\Big) ~+~ \frac12 \cdot Ent\Big(g~|~x_1 = 1,x_2,...,x_n\Big)
\]

\noi As before, let $g_T = \E\Big(h~|~T\Big)$, for a subset $T \subseteq [n]$. Note that if $1 \in T$, then $\E\Big(g_T~|~\{1\}\Big) = \E\Big(h~|~\{1\}\Big)$.

\noi Hence
\[
\E_{T,1 \in T} ~\bigg(Ent\Big(h~|~T\Big) - Ent\Big(h~|~\{1\}\Big) \bigg) ~~=~~ \E_{T,1 \in T} ~\bigg(Ent\Big(g_T\Big) - Ent\Big(g_T~|~\{1\}\Big) \bigg) ~~=
\]
\[
\frac12 ~\cdot \E_{T,1 \in T} Ent\Big(g_T~|~x_1 = 0,x_2,...,x_n\Big) ~+~ \frac12 ~\cdot \E_{T,1 \in T} Ent\Big(g_T~|~x_1 = 1,x_2,...,x_n\Big)
\]

\noi We will prove the lemma by showing that, for a sufficiently large absolute constant $L$, hold both
\beqn
\label{x1=0}
\E_{T,1 \in T} Ent\Big(g_T~|~x_1 = 0,x_2,...,x_n\Big) \quad \le \quad L \cdot \l_1 \cdot \gamma
\eeqn
and
\beqn
\label{x1=1}
\E_{T,1 \in T} Ent\Big(g_T~|~x_1 = 1,x_2,...,x_n\Big) ~~ \le ~~ L \cdot \bigg(\l_1 \cdot \gamma ~+~ \gamma^2 \ln\(\frac{1}{\gamma}\) \bigg)
\eeqn

\

\subsubsection*{Proof of (\ref{x1=0})}

\noi Fix a subset $T \subseteq [n]$, with $1 \in T$. Recall that $g_T = \sum_{S \subseteq T} \widehat{h}(S) \cdot W_S$, and hence
\[
\E \Big(g_T~|~x_1 = 0,x_2,...,x_n\Big) = \sum_{R \subseteq T \setminus \{1\}}  \Big(\widehat{h}(R) ~+~ \widehat{h}(R \cup \{1\}) \Big) \cdot W_R
\]
In particular,
\[
\E \Big(g_T~|~x_1 = 0\Big) ~~=~~ \widehat{h}(0) ~+~ \widehat{h}(\{1\}) ~~=~~ \widehat{f}(0) ~+~ \tau^{1/2} \cdot \widehat{f}(\{1\}) ~~\ge~~ \E f
\]

\noi Applying (\ref{LS-F}), we have, for a sufficiently large constant $L_1$,
\[
Ent\Big(g_T~|~x_1 = 0,x_2,...,x_n\Big) ~~\le~~ \frac{2}{\ln 2} \cdot \frac{1}{\E f} \cdot \sum_{R \subseteq T \setminus \{1\}}  |R| \cdot \Big(\widehat{h}(R) ~+~ \widehat{h}(R \cup \{1\}) \Big)^2 ~~\le
\]
\[
L_1 \cdot \sum_{R \subseteq T \setminus \{1\}}  |R| \cdot \Big(\widehat{h}^2(R) ~+~ \widehat{h}^2(R \cup \{1\})\Big)
\]

\noi Averaging over $T$, we have
\[
\E_{T,1 \in T} ~Ent\Big(g_T~|~x_1 = 0,x_2,...,x_n\Big) ~\le~ L_1 \cdot \bigg(\sum_{R, 1 \not \in R} ~|R| \l_1^{|R|} \widehat{h}^2(R) ~+~ \sum_{R, 1 \not \in R} ~|R| \l_1^{|R|} \widehat{h}^2(R \cup \{1\}) \bigg)
\]

\noi Using the fact that $|\widehat{h}(S)| \le |\widehat{f}(S)|$ for all $S \subseteq [n]$, and applying (\ref{fl>2}) and (\ref{fl+others}), we have, for a sufficiently large constant $L_2$,
\[
\sum_{R, 1 \not \in R} ~|R| \l_1^{|R|} \widehat{h}^2(R) ~~\le~~ L_2 \cdot \bigg( \l_1 \cdot \gamma^2 \ln\(\frac{1}{\gamma}\) ~+~ \l^2_1 \cdot \gamma \bigg)
\]
and
\[
\sum_{R, 1 \not \in R} ~|R| \l_1^{|R|} \widehat{h}^2(R \cup \{1\}) ~~\le~~ L_2 \cdot \l_1 \cdot \gamma
\]

\noi Summing up, this gives (\ref{x1=0}).

\subsubsection*{Proof of (\ref{x1=1})}

\noi Similarly to the above,
\[
\E \Big(g_T~|~x_1 = 1,x_2,...,x_n\Big) = \sum_{R \subseteq T \setminus \{1\}}  \Big(\widehat{h}(R) ~-~ \widehat{h}(R \cup \{1\}) \Big) \cdot W_R
\]
Which means that
\[
\E \Big(g_T~|~x_1 = 1\Big) ~~=~~ \widehat{h}(0) ~-~ \widehat{h}(\{1\}) ~~=~~ \widehat{f}(0) ~-~ \tau^{1/2} \cdot \widehat{f}(\{1\}) ~~=~~ \E f \cdot \Big(1 ~-~ \tau^{1/2} \cdot (1-\alpha)\Big)
\]

\noi Recall that $\tau^{1/2} = 1$ if $\alpha \ge \l$ and $\tau^{1/2} = \frac{1-\l}{1-\alpha}$ otherwise. In both cases, note that we have $\E \Big(g_T~|~x_1 = 1\Big) \ge \l \cdot \E f$.

\noi Applying (\ref{LS-F}), and averaging over $T$, we have, for a sufficiently large constant $L_1$,
\[
\E_{T,1 \in T} ~Ent\Big(g_T~|~x_1 = 1,x_2,...,x_n\Big) ~~\le~~ L_1 \cdot \frac{1}{\l} \cdot \sum_{R, 1 \not \in R}  |R| \l_1^{|R|} \cdot \Big(\widehat{h}(R) ~-~ \widehat{h}(R \cup \{1\}) \Big)^2
\]

\noi Let $g = \E \Big(h~|~x_1 = 1,x_2,...,x_n\Big)$. Then $g = \sum_{R \subseteq [n], 1 \not \in R} ~\Big(\widehat{h}(R) ~-~ \widehat{h}(R \cup \{1\}) \Big) \cdot W_R$. Hence
\beqn
\label{x1=1,ent}
\E_{T, 1 \in T} ~Ent\Big(g_T~|~x_1 = 1,x_2,...,x_n\Big) ~~\le~~ L_1 \cdot \frac{1}{\l} \cdot \sum_{R, 1 \not \in R}  |R| \l_1^{|R|} \cdot \widehat{g}^2(R)
\eeqn

\noi Consider the function $g$. Since $h = T_{\e_{\t}} f$, we have
\[
g ~~=~~ \e_{\t} \cdot T_{\e_{\t}} \bigg(\E \Big(f~|~x_1 = 0,x_2,...,x_n\Big)\bigg) ~+~ \Big(1 ~-~ \e_{\t}\Big) \cdot T_{\e_{\t}} \bigg(\E \Big(f~|~x_1 = 1,x_2,...,x_n\Big)\bigg)
\]

\noi For $i = 0, 1$, let $f_i = \E \Big(f~|~x_1 = i,x_2,...,x_n\Big)$, and let $t_i = T_{\e_{\t}} f_i$. Note that for $i = 0, 1$ and for any $R$, $1 \not \in R$, holds $|\widehat{t_i}(R)| ~\le~ |\widehat{f_i}(R)|$.

\noi Therefore, since $g = \e_{\t} \cdot t_0 + \Big(1 - \e_{\t}\Big) \cdot t_1$, we have, for any $R$, $1 \not \in R$ that
\[
\widehat{g}^2(R) ~~\le~~ \e_{\t} \cdot \widehat{t_0}^2(R) ~+~ \Big(1 ~-~ \e_{\t}\Big) \cdot \widehat{t_1}^2(R) ~~\le~~ \e_{\t} \cdot \widehat{f_0}^2(R) ~+~ \Big(1 ~-~ \e_{\t}\Big) \cdot \widehat{f_1}^2(R)
\]

\noi Hence,
\beqn
\label{x1=1,sum}
\sum_{R, 1 \not \in R}  |R| \l_1^{|R|} \cdot \widehat{g}^2(R) ~~\le~~  \e_{\t} \cdot \sum_{R, 1 \not \in R}  |R| \l_1^{|R|} \widehat{f_0}^2(R) ~+~ \Big(1 - \e_{\t}\Big) \cdot \sum_{R, 1 \not \in R}  |R| \l_1^{|R|} \widehat{f_1}^2(R)
\eeqn

\noi Exactly as above, we have the following upper bound for the first summand: For a sufficiently large constant $L_2$ holds
\[
\sum_{R, 1 \not \in R}  |R| \l_1^{|R|} \widehat{f_0}^2(R) ~~=~~ \sum_{R, 1 \not \in R}  |R| \l_1^{|R|} \Big(\widehat{f}(R) ~+~ \widehat{f}(R \cup \{1\}) \Big)^2 ~~\le~~ L_2 \cdot \l_1 \cdot \gamma
\]

\noi Consider the second summand. The function $f_1$ is a boolean function, whose expectation equals $\widehat{f}(0) - \widehat{f}(\{1\}) = \alpha \cdot \E f \le \alpha$. Similarly, $\E f_1^2 = \E f_1 \le \alpha$.

\noi We now apply the inequality of \cite{Talagrand}, which states that

\noi {\it For a boolean function $g:~\{0,1\}^m \rarrow \{0,1\}$ with expectation $\mu \le 1/2$ holds $\sum_{k=1}^m \widehat{g}^2(\{k\}) \le L_3 \cdot \mu^2 \cdot \ln\(1/\mu\)$, for a sufficiently large absolute constant $L_3$.}

\noi In our case, this implies $\sum_{k=2}^n \widehat{f_1}^2\Big(\{k\}\Big) \le L_3 \cdot \alpha^2 \cdot \ln\(\frac{1}{\alpha}\)$, for a sufficiently large constant $L_3$.

\noi This means that, for a sufficiently large constant $L_4$, we can upperbound the second summand in (\ref{x1=1,sum}) by
\[
\sum_{R, 1 \not \in R}  |R| \l_1^{|R|} \widehat{f_1}^2(R) ~~\le~~ L_4 \cdot \bigg(\l_1 \cdot \alpha^2 \ln\(\frac{1}{\alpha}\) ~+~ \l^2_1 \cdot \alpha \bigg)
\]

\noi Recall that for $\alpha < \l$, we have $\e_{\t} = \frac{1 - \tau^{1/2}}{2} =  \frac{1 - (1-\l)/(1-\alpha)}{2} \le L_5 \cdot \l$, for an absolute constant $L_5$; and that for $\alpha \ge \l$, we have $\e_{\t} = 0$. Plugging these estimates into (\ref{x1=1,sum}), we have
\[
\sum_{R, 1 \not \in R}  |R| \l_1^{|R|} \cdot \widehat{g}^2(R) ~~\le~~ L_2 \cdot L_5 \cdot \l \cdot \l_1 \cdot \gamma ~+~ L_4 \cdot \bigg(\l_1 \cdot \alpha^2 \ln\(\frac{1}{\alpha}\) ~+~ \l^2_1 \cdot \alpha \bigg)
\]
And hence, coming back to (\ref{x1=1,ent}), and recalling that $\l = \tau \cdot \l_1$, we have, for sufficiently large absolute constants $L$, $L'$, that
\[
\E_{T, 1 \in T} ~Ent\Big(g_T~|~x_1 = 1,x_2,...,x_n\Big) ~~\le~~ L' \cdot \bigg(\l_1 \cdot \gamma ~+~ \alpha^2 \ln\(\frac{1}{\alpha}\) ~+~ \l_1 \cdot \alpha \bigg) ~~\le
\]
\[
L \cdot \bigg(\l_1 \cdot \gamma ~+~ \gamma^2 \ln\(\frac{1}{\gamma}\) \bigg)
\]

\noi This completes the proof of (\ref{x1=1}), of Lemma~\ref{lem:1-in-B}, and of the first claim of the theorem.

\eprf

\subsection{Proof of the second claim of the theorem}

\noi First, note that if $f$ is balanced, that is $\E f = \frac12$, then so is $1-f$, and the second claim of the theorem follows immediately from the first claim.

\noi If $\E f \not = \frac12$, some additional work is required. We only sketch the argument below, since it is very similar to the proof of the first claim.

\noi Applying Theorem~\ref{thm:main} to the function $1-f$ gives (cf. (\ref{stab}))
\[
Ent\Big(T_{\e} (1-f)\Big) = Ent\Big(T_{\e_1} (1-h)\Big) \le \l_1 \cdot \E_{T,1 \in T} \bigg(Ent\Big((1-h)~|~T\Big) - Ent\Big((1-h)~|~\{1\}\Big) \bigg)  ~+
\]
\beqn
\label{stab-2}
\(1 - \l_1\) \cdot \E_{T,1 \not \in T} ~Ent\Big((1-h)~|~T\Big) ~~+~~\E (1-h) \cdot \phi\Bigg(Ent\(\frac{1-h}{\E (1-h)}~ \Big |~\{1\}\),~\e_1\Bigg)
\eeqn

\noi As in the proof of the first claim, we upperbound each of the three summands on the RHS of (\ref{stab-2}) separately.

\noi Repeating the argument, with the necessary (minor) differences, leads to the same first two bounds:

\begin{itemize}

\item
\[
\E_{T,1 \in T} ~\bigg(Ent\Big((1-h)~|~T\Big) - Ent\Big((1-h)~|~\{1\}\Big) \bigg) \quad \le \quad O\Bigg(\l_1 \cdot \gamma ~+~ \gamma^2 \ln\(\frac{1}{\gamma}\)\Bigg)
\]

\item
\[
\E_{T,1 \not \in T} ~Ent\Big((1-h)~|~T\Big) \quad \le \quad O\Bigg(\l^2_1 \cdot \gamma ~+~ \l_1 \cdot \gamma^2 \ln\(\frac{1}{\gamma}\)\Bigg)
\]

\end{itemize}

\noi Indeed, this should not be surprising since, roughly speaking, these two bounds for $h$ are obtained by analysing the behavior of (the squares of) its non-trivial Fourier coefficients, and this is the same for $h$ and for $1-h$.

\noi As to the third summand, we will follow the argument in the proof of Lemma~\ref{lem:marginal-1}.

\noi Let $g = \E\(\frac{1-f}{\E (1-f)}~\Big | ~\{1\}\)$. This is a function on a $2$-point space $\{0,1\}$, with $g(0) = \rho$ and $g(1) = 2 - \rho$, where $\rho = 1 + \frac{(1-\alpha)(1 - 2\beta)}{1 + 2\beta}$.

\noi Note that $\rho \le 2 - c \cdot \gamma$, for some absolute constant $c > 0$. Hence, proceeding as in the proof of Lemma~5.3, gives
\[
\phi\Bigg(Ent\(\frac{1-h}{\E (1-h)}~ \Big |~\{1\}\),~\e_1\Bigg) ~~=~~  Ent\Big(T_{\e} g \Big) ~~\le~~ \Big(1 - H_2(\e)\Big) - c' \cdot \l \cdot \gamma,
\]
for an absolute constant $c' > 0$.

\noi Next, recall that in the proof of Lemma~\ref{lem:marginal-1} we show
$
\phi\(Ent\(\frac{h}{\E h}~ \Big |~\{1\}\),~\e_1\) ~\le~ \Big(1 - H_2(\e)\Big) - c_1 \cdot \l \cdot \alpha
$
for an absolute constant $c_1 > 0$.

\noi Hence
\[
\E h \cdot \phi\Bigg(Ent\(\frac{h}{\E h}~ \Big |~\{1\}\),~\e_1\Bigg)
~+~ \E (1-h) \cdot \phi\Bigg(Ent\(\frac{1-h}{\E (1-h)}~ \Big |~\{1\}\),~\e_1\Bigg) ~~\le~~
\]
\[
\Big(1 - H_2(\e)\Big) - c_2 \cdot \l \cdot \gamma,
\]
for an absolute constant $c_2 > 0$.

\noi We can now complete the proof of the second claim of the theorem.

\noi Combining all bounds on the right hand sides of (\ref{stab}) and of (\ref{stab-2}) above gives
\[
Ent\Big(T_{\e} f\Big) ~+~ Ent\Big(T_{\e} (1-f) \Big)  ~~\le~~ \Big(1 - H_2(\e)\Big)  ~-~ \Omega\Big(\l \cdot \gamma \Big) ~+~ o_{\l, \gamma \rarrow 0} \Big(\l \cdot \gamma \Big)
\]

\noi Since $\l, \gamma ~\le~ O(\delta)$, this implies that for a sufficiently small $\delta > 0$ holds
\[
Ent\Big(T_{\e} f\Big) ~+~ Ent\Big(T_{\e} (1-f) \Big)  ~~\le~~ 1 - H_2(\e)
\]

\eprf

\section{Remaining proofs}
\label{sec:remaining}

\subsection{Proof of Lemma~\ref{lem:H-entr}}
We have, for a boolean function $f$:
\[
I\Big(f(X);Y\Big) ~~=~~ H\Big(f(X)\Big) ~-~ H\Big(f(X) | Y\Big) ~~=~~ H\Big(f(X)\Big) ~-~ \E_y H\Big(f(X) | Y = y\Big) ~~=
\]
\[
H_2\Big(\E f\Big) ~-~ \E_y H_2\Big(\(T_{\e} f\)(y)\Big)
\]
We have $H_2\(\E f\) = \E f \log \frac{1}{\E f} + (1-\E f) \log \frac{1}{1 - \E f}$.

\noi We also have (all the logarithms are binary)
\[
\E_y H_2\Big(\(T_{\e} f\)(y)\Big) ~~=~~ \E_y \( \(T_{\e} f\)(y) \log \frac{1}{\(T_{\e} f\)(y)} ~+~ \Big(1 - \(T_{\e} f\)(y)\Big) \log \frac{1}{ 1- \(T_{\e} f\)(y)} \)  ~~=
\]
\[
- \bigg(Ent\Big(T_{\e} f\Big) ~+~ \E T_{\e} f \log \E T_{\e} f\bigg) ~~-~~ \bigg( Ent\Big(T_{\e} (1-f)\Big) ~+~ \E T_{\e} (1-f) \log \E T_{\e} (1-f) \bigg) ~~=
\]
\[
- \bigg(Ent\Big(T_{\e} f\Big) + Ent\Big(T_{\e} (1-f)\Big) \bigg) ~+~ \E f \log \frac{1}{\E f} ~+~ (1-\E f) \log \frac{1}{1 - \E f}
\]
In the last step we have used the fact $\E T_{\e} g ~=~ \E g$ for any function $g$. The claim of the lemma follows.
\eprf

\subsection{Proof of Corollary~\ref{cor:info-app}}

\noi Applying Corollary~\ref{cor:streamline} to the functions $f$ and $1-f$, we obtain, by Lemma~\ref{lem:H-entr}:
\[
I\Big(f(X);Y\Big) ~~=~~ Ent\Big(T_{\e} f\Big) ~+~ Ent\Big(T_{\e} (1-f)\Big) ~\le~
\E_T \bigg(Ent\Big(f~|~T\Big) ~+~ Ent\Big((1-f)~|~T\Big)\bigg)
\]

\noi To conclude the proof of the corollary, it suffices to show that for any $T \subseteq [n]$ holds
\[
Ent\Big(f~|~T\Big) ~+~ Ent\Big((1-f)~|~T\Big) ~~=~~ I\Big(f(X);~\{X_i\}_{i \in T}\Big)
\]

\noi To see this, we proceed exactly as in the proof of Lemma~\ref{lem:H-entr}, observing that, by definition,
\[
Pr\Big\{f(X) = 1 ~\Big |~ \{X_i\}_{i \in T}\Big\} \quad = \quad \E\Big(f~|~T\Big)
\]

\noi Here we interpret both sides as functions of $\{x_i\}$, $i \in T$.

\eprf

\subsection{Proof of Theorem~\ref{thm:sMGL}}
\label{sec:sMGL}

\noi The proof of this theorem is very similar to that of Theorem~\ref{thm:main} and uses the notation and some of the results from that proof.

\noi As in the proof of Lemma~\ref{lem:noisy_ts}, our starting point is the chain rule for noisy entropy (\ref{chain-rule}), which states that for any permutation $\sigma \in S_n$ the noisy entropy $Ent\Big(T_{\e} f\Big)$ is bounded from above by
\[
\sum_{i=1}^n \phi\Bigg(Ent\Big(f~|~\{\sigma(i)\}\Big) ~~+~~ I_{T_{\e_{\left\{\sigma(1),\ldots,\sigma(i-1)\right\}}} f}~\bigg(\Big\{\sigma(1),\ldots,\sigma(i-1)\Big\},~\sigma(i)\bigg)\Bigg)
\]

\noi Averaging over $\sigma \in S_n$ and using transitivity of action of the symmetric group and concavity of $\phi$, this is at most
\[
\sum_{k=0}^{n-1} \phi\(\E_{i \in [n]} ~Ent\Big(f~|~i\Big) ~~+~~ b_k\) \quad \mbox{where} \quad b_k \quad = \quad \E_{A, m} ~T_{\e_A f}~\Big(A,m\Big)
\]
and the expectation is over all $A \subseteq [n]$ of cardinality $k$ and $m \not \in A$. (In particular, we set $b_0 = 0$). Using the concavity of $\phi$ again, this is at most
\[
n \cdot \phi\(\E_{i \in [n]} ~Ent\Big(f~|~i\Big) ~~+~~ \frac{1}{n} \cdot \sum_{k=0}^{n-1} b_k\)
\]

\noi The analysis in the proof of Theorem~\ref{thm:main} shows that if $T$ is a random subset of $[n]$ generated by sampling each element $i \in [n]$ independently with probability $\l$ then
\[
\sum_{k=0}^{n-1} b_k ~~ = ~~ \frac{1}{\l} \cdot \sum_{s=1}^{n-1} w_s \cdot t_s ~~=~~ \frac{1}{\l} \cdot \E_T \bigg(Ent\Big(f~|~T\Big) ~-~ \sum_{i \in T} Ent\Big(f~|~i\Big) \bigg) ~~=~~
\]
\[
\frac{1}{\l} \cdot \E_T Ent\Big(f~|~T\Big) ~-~  n \cdot \E_{i \in [n]} ~Ent\Big(f~|~i\Big)
\]
Substituting and simplifying, we get, setting $t = \l n$,
\[
Ent\Big(T_{\e} f\Big) \quad \le \quad n \cdot \phi\(\frac{\E_T Ent\Big(f~|~T\Big)}{t}\)
\]
which is the claim of the theorem.
\eprf

\subsubsection{Proof of (\ref{sMGL-infor})}


\noi Let $f$ be the distribution of $X$ multiplied by $2^n$. Then $\E f = 1$, and Theorem~\ref{thm:sMGL} can be applied.

\noi By Section~\ref{subsubsec:connect} and (\ref{connect-dist}), we have $Ent\Big(T_{\e} f\Big) ~=~ n - H\Big(X \oplus Z\Big)$ and $Ent\Big(f~|~T\Big)~=~|T| - H\Big(\{X_i\}_{i\in T}\Big)$.

\noi We also recall $\phi\Big(x,\e\Big) ~=~ 1 - H_2\Big(\e + (1-2\e) \cdot H_2^{-1}(1-x)\Big)$.

\noi Substituting in the claim of the theorem, and simplifying, gives
\[
H\Big(X \oplus Z\Big) \quad \ge \quad n \cdot H_2\(\e ~+~ (1-2\e) \cdot H_2^{-1}\(\frac{\E_{T} H\Big(\{X_i\}_{i\in T}\Big)}{t}\)\)
\]
which is the claim of (\ref{sMGL-infor}).

\eprf

\subsection{Proof of Theorem~\ref{thm:KC-bsn}}

\noi Let $\delta$ be the constant in the theorem. We will assume that $\delta$ is sufficiently small.

\noi Let $\e$ be a noise parameter, such that $(1-2\e)^2 \le \delta$. Denote $\l = (1-2\e)^2$.

\noi It is known (see \cite{Kumar-Courtade}) that for any boolean function $f$ holds $I\(f(X);Y\) ~\le~ \l \cdot H_2\(\E f\)$. This immediately implies the validity of Conjecture~\ref{cnj:1} for boolean functions with expectation lying in $[0,c] ~\cup~ [1-c,1]$, for some absolute constant $0 < c < 1/2$.

\noi In addition, we may assume, by symmetry, that $\E f \le 1/2$. Combining these two observations, it remains to consider the case
\beqn
\label{exp-bal}
c ~~\le~~ \E f ~~\le~~ 1/2
\eeqn

\noi Let $f$ be a boolean function satisfying (\ref{exp-bal}) with $I\(f(X);Y\) \ge  1 - H_2(\e)$. This is the same as $Ent\(T_{\e} f\) + Ent\(T_{\e} (1-f)\) \ge  1 - H_2(\e)$.

\noi At this point, we need a technical lemma.

\lem
\label{lem:first level}
For any nonnegative non-zero function $f$ holds\footnote{Asymptotic notation hides absolute constants independent of the remaining parameters.}
\[
Ent\Big(T_{\e} f\Big) ~~ \le ~~ \frac{1}{\E f} \cdot \(\frac{1}{2\ln 2} \cdot \sum_{k=1}^n \widehat{f}^2(\{k\})\) \cdot \l  ~+~  O_{\l \rarrow 0}\(\frac{\E f^2}{\E f} \cdot \l^{4/3}\) ~+~  O_{\l \rarrow 0}\(\frac{\E^2 f^2}{\E^3 f} \cdot \l^2\)
\]
\elem

\noi We will now proceed with the proof of the theorem, and prove the lemma below.

\noi Applying Lemma~\ref{lem:first level} to functions $f$ and $1-f$ and taking into account (\ref{exp-bal}) gives
\[
Ent\Big(T_{\e} f\Big) ~+~ Ent\Big(T_{\e} (1-f)\Big) ~~\le~~ \frac{1}{\E f \Big( 1- \E f\Big)} \cdot \(\frac{1}{2\ln 2} \cdot \sum_{k=1}^n \widehat{f}^2(\{k\})\) \cdot \l  ~+~ O_{\l \rarrow 0}\Big(\l^{4/3}\Big)
\]

\noi Combining the two inequalities for $Ent\(T_{\e} f\) + Ent\(T_{\e} (1-f)\)$, recalling $1 - H_2(\e) ~\ge~ \frac{\l}{2 \ln 2}$, and using (\ref{exp-bal}), we get
\[
\sum_{k=1}^n \widehat{f}^2(\{k\}) \quad \ge \quad \E f \cdot \Big(1 - \E f\Big) ~-~ O\(\l^{1/3}\)
\]
For a boolean function $f:\H \rarrow \{0,1\}$ holds $\E f^2 = \E f$, and consequently $\sum_{S \not = \emptyset} \widehat{f}^2(S) ~=~ \E f ( 1 - \E f)$. Hence we have
\[
\sum_{|S| \ge 2} \widehat{f}^2(S) \le O\(\l^{1/3}\)
\]

\noi Let $g = 2 f - 1$. Then $g:~\H \rarrow \{-1,1\}$, and $\sum_{|S| \ge 2} \widehat{g}^2(S) = 4 \cdot \sum_{|S| \ge 2} \widehat{f}^2(S) \le O\(\l^{1/3}\)$.

\noi Note also that (\ref{exp-bal}) implies $2c - 1 \le \widehat{g}(0) = 2\E g - 1 \le 0$.

\noi Hence, assuming $\l$ is sufficiently small, Theorem~\ref{thm:JOW} implies that $|\E g| \le O\(\l^{1/3} \cdot \sqrt{\ln \frac{1}{\l}}\)$, and there exists an index $1 \le k \le n$ such that $\widehat{g}^2(\{k\}) \ge 1 - O\(\l^{1/3}\)$.

\noi This means that
\[
\frac12 ~-~ O\(\l^{1/3} \cdot \sqrt{\ln \frac{1}{\l}}\) ~~\le~~ \E f ~~\le~~ \frac12 \quad \mbox{and} \quad |\widehat{f}(\{k\})| ~~\ge~~ \frac12 ~-~ O\(\l^{1/3}\)
\]

\noi If $\l$ is sufficiently small, $f$ satisfies the conditions of Theorem~\ref{thm:KC-ent-sn}. By the second claim of this theorem,
\[
Ent\(T_{\e} f\) ~+~ Ent\(T_{\e} (1-f)\) ~~\le~~  1 ~-~ H_2(\e),
\]
completing the proof of Theorem~\ref{thm:KC-bsn}.

\eprf

\subsubsection{Proof of Lemma~\ref{lem:first level}}

\noi The argument below is a slight extension of an argument in \cite{OSW}.

\noi In the following, we may and will assume, by homogeneity, that $\E f = 1$.

\noi Let us introduce some notation. For $x \in \H$, let $x^c$ be the complement of $x$, that is the element of $\H$ with $x^c_i = 1 - x_i$ for all $1 \le i \le n$.

\noi For a nonnegative function $g$ on $\H$, let $g_0$ be the 'even' part of $g$ defined by $g_0(x) ~=~ \(g(x) + g\(x^c\)\)/2$, and let $g_1 ~=~ g - g_0$ be the 'odd' part of $g$. By definition, $g_0(x) ~=~ g_0\(x^c\)$ and $g_1(x) ~=~ -g_0\(x^c\)$. Note also that $|g_1| \le g_0$.

\noi We will need the following well-known (and easy to verify) fact:
\[
g_0 ~~=~~ \sum_{|S| ~even} \widehat{g}(S) \cdot W_S \quad \mbox{and} \quad g_1 ~~=~~ \sum_{|S| ~odd} \widehat{g}(S) \cdot W_S
\]

\noi We start with an auxiliary claim.
\lem
\label{lem:aux}
For a function $g$ with $\E g = 1$ holds
\[
Ent\(g\) ~~=~~ Ent\Big(g_0\Big) ~+~ \E_x ~g_0(x) \cdot \Bigg(1 - H_2\(\frac{1 - |g_1(x)|/g_0(x)}{2}\)\Bigg)
\]
Here for $x$ such that $g_0(x) = g_1(x) = 0$, the expression $g_0(x) \cdot \(1 - H_2\(\frac{1 - |g_1(x)|/g_0(x)}{2}\)\)$ is interpreted as $0$.
\elem

\prf
We have
\[
Ent\Big(g\Big) \quad = \quad \E_x g(x) \log g(x) \quad = \frac12 \cdot \E_x \Big( g(x) \log g(x) ~+~ g\(x^c\) \log \(x^c\) \Big) \quad =
\]
\[
\frac12 \cdot \E_x \bigg(\Big(g_0(x) + g_1(x) \Big) \cdot \log \Big(g_0(x) + g_1(x) \Big) ~~+~~ \Big(g_0(x) - g_1(x) \Big) \cdot \log \Big(g_0(x) - g_1(x) \Big)\bigg)
\]
It is easy to verify that for any $0 \le b \le a$ holds \\
$1/2 \cdot \Big((a+b) \log(a+b) + (a-b) \log(a-b)\Big) ~=~ a \log a + a \cdot \(1 - H_2\(\frac{1 - b/a}{2}\)\)$, where the last expression should be interpreted as $0$ for $a = b = 0$.

\noi Using this identity with $a = g_0(x)$ and $b = g_1(x)$ gives the claim of the lemma.
\eprf

\noi Next, as in \cite{OSW}, we upper bound $1 - H_2\(\frac{1-x}{2}\)$ by $\frac{1}{2 \ln 2} \cdot x^2 + \(1 - \frac{1}{2 \ln 2}\) \cdot x^4$.

\noi Substituting this bound in the claim of Lemma~\ref{lem:aux} gives
\[
Ent\(g\) ~~\le~~ Ent\Big(g_0\Big) ~~+~~ \frac{1}{2 \ln 2} \cdot \E_x ~\frac{g^2_1(x)}{g_0(x)} ~~+~~ \(1 - \frac{1}{2 \ln 2}\) \cdot \E_x ~\frac{g^4_1(x)}{g^3_0(x)}
\]

\noi Let $g = T_{\e} f$. It is easy to verify $\Big(T_{\e} f\Big)_i = T_{\e} \Big(f_i\Big)$ for any function $f$ and $i = 0, 1$. Consequently, we get the bound
\[
Ent\Big(T_{\e} f\Big) ~~\le~~ Ent\Big(T_{\e} f_0\Big) ~+~ \frac{1}{2 \ln 2} \cdot \E_x ~\frac{\Big(T_{\e} f_1(x)\Big)^2}{T_{\e} f_0(x)} ~+~ \(1 - \frac{1}{2 \ln 2}\) \cdot \E_x ~\frac{\Big(T_{\e} f_1(x)\Big)^4}{\Big(T_{\e} f_0(x)\Big)^3}
\]

\noi We upperbound each of the summands on the RHS separately.

\begin{enumerate}

\item
The first summand. Note that $\E T_{\e} f_0 = \E f_0 = \E f = 1$. Recall also (see e.g., \cite{OSW}) that for any function $g$ on $\H$ holds $T_{\e} g ~=~ \sum_S \l^{|S|/2} \widehat{g}(S) \cdot W_S$.

\noi Hence, by Lemma~\ref{lem:LS-1},
\[
Ent\Big(T_{\e} f_0\Big) \quad \le \quad O\bigg(\sum_S |S| \cdot \widehat{T_{\e} f_0}^2(S) \bigg) \quad = \quad O\bigg(\sum_S |S| \l^{|S|} \widehat{f_0}^2(S) \bigg) \quad =
\]
\[
O\bigg(\sum_{|S|~even} ~|S| \l^{|S|} \widehat{f}^2(S) \bigg) \quad = \quad O\Big(\E f^2 \cdot \l^2\Big)
\]

\item
The second summand.

\noi First, we argue that $T_{\e} f_0$ is bounded away from $0$ with high probability. Recall that $\E T_{\e} f_0 ~=~ 1$, and note that $Var\Big(T_{\e} f_0\Big) ~=~ \sum_{S \not = 0} \widehat{T_{\e} f_0}^2(S) ~=~ O\Big(\l^2 \cdot \E f^2\Big)$. Hence, by Chebyshev's inequality, for any $0 \le \alpha < 1$ holds
\beqn
\label{cheb}
Pr\Big\{T_{\e} f_0 \le \alpha\Big\} ~~\le~~ O\(\frac{\l^2 \cdot \E f^2}{(1-\alpha)^2}\)
\eeqn

\noi Second, recall that for any $x$ holds $|T_{\e} f_1(x)| \le T_{\e} f_0(x)$, and hence $\frac{\(T_{\e} f_1(x)\)^2}{T_{\e} f_0(x)} ~\le~ T_{\e} f_0(x)$.

\noi Therefore, taking $\alpha ~=~ 1 - \l^{1/3}$ in (\ref{cheb}), we have $\E_x \frac{\(T_{\e} f_1(x)\)^2}{T_{\e} f_0(x)}$ bounded from above by
\[
Pr\Big\{f_0 ~\le~ 1 - \l^{1/3} \Big\} \cdot \alpha  ~~+~~ \Big(1 + O\(\l^{1/3}\)\Big) \cdot \E_x \Big(T_{\e} f_1(x)\Big)^2
\]

\noi Recalling that $T_{\e} f_1 ~=~ \sum_{|S|~odd} \l^{|S|/2} \widehat{f}(S) \cdot W_S$, this is at most

\[
O\Big(\E f^2 \cdot \l^{4/3}\Big) ~~+~~  \Big(1 + O\(\l^{1/3}\)\Big) \cdot \(\(\sum_{k=1}^n \widehat{f}^2(\{k\})\) \cdot \l ~+~ O\Big(\E f^2 \cdot \l^3 \Big)\) \quad =
\]
\[
\(\sum_{k=1}^n \widehat{f}^2(\{k\})\) \cdot \l ~+~ O\Big(\E f^2  \cdot \l^{4/3} \Big)
\]

\item
The third summand. Note that $\E f_1 = 0$. Hence, as in Lemma~1 in \cite{OSW} (where the requirement on $f$ to be boolean does not seem to be necessary) we have, for a sufficiently small $\l$, that
\[
\E_x \Big(T_{\e} f_1(x)\Big)^4 ~~\le~~ O\(\(\E_x f^2_1(x)\)^2 \cdot \l^2\) ~=~ O\({\E}^2 f^2 \cdot \l^2 \)
\]
We can now upperbound the third summand using the Chebyshev inequality, as above. Taking $\alpha = 1/2$ in (\ref{cheb}) upperbounds $\E_x \frac{\(T_{\e} f_1(x)\)^4}{\(T_{\e} f_0(x)\)^3}$ by
\[
O\(\E f^2 \cdot \l^2\)  \cdot \alpha ~+~ O\({\E}^2 f^2 \cdot \l^2 \) ~~=~~ O\(\E f^2 \cdot \l^2\)  ~+~ O\({\E}^2 f^2 \cdot \l^2 \)
\]

\end{enumerate}

\noi Combining these estimates leads to the claim of Lemma~\ref{lem:first level}.
\eprf

\eprf

\section*{Acknowledgments}
We are grateful to Yuval Kochman, Or Ordentlich, and Yury Polyanskiy for many very helpful conversations and valuable remarks. We also thank Venkat Chandar for valuable remarks.

\end{document}